\documentclass[twocolumn, nofootinbib, aps, prd, 10pt]{revtex4-1}
\usepackage{graphicx}
\usepackage{float}
\usepackage{url}
\usepackage{color}
\usepackage{rotating}
\usepackage{amssymb,amsmath}
\usepackage{ulem} 
\usepackage{mathtools} 

\usepackage{xspace}
\newcommand{\sect}[1]{Sec.~\ref{#1}\xspace}

\newcommand{\fig}[1]{Fig.~\ref{#1}\xspace}
\newcommand{\eq}[1]{Eq.~(\ref{#1})\xspace}
\newcommand{\tab}[1]{Table~\ref{#1}\xspace}

\newcommand{\degr}{\hbox{$^\circ$}}

\newcommand{\fsky}{f_{\rm sky}}

\newcommand{\alm}{a_{\ell m}}

\newcommand{\Cl}{C_\ell}

\renewcommand{\vec}[1]{\boldmath{#1}}

\newcommand{\ellmax}  {\ell_{\mathrm{max}}}

\newcommand{\nside}{\textsc{Nside}}

\newcommand{\Var}{\rm Var}

\newcommand{\nhat}{{\bf \hat n}}
\newcommand{\healpix}{\rm {\tt HEALPix}\xspace}

\newcommand{\quadrupole}{C_{2}}
\newcommand{\octopole}{C_{3}}
\newcommand{\SQO}{S_{\rm QO}}
\newcommand{\ALV}{A_{\rm LV}}
\newcommand{\svar}{\sigma_{16}^2}
\newcommand{\Shalf}{S_{1/2}}
\newcommand{\Rparity}{R_{27}}
\newcommand{\ctpi}{C(\pi)}
\newcommand{\PIS}{Planck XVI (I\&S)\xspace}

\everymath{\displaystyle}

\newcommand{\be}{\begin{equation}}
\newcommand{\ee}{\end{equation}}

\begin{document}

\title{Covariance of CMB anomalies}

\author{Jessica Muir}
\affiliation{Department of Physics, University of Michigan, 
  450 Church St, Ann Arbor, MI 48109-1040, \\
  Leinweber Center for Theoretical Physics, University of Michigan, 
  450 Church St, Ann Arbor, MI 48109-1040}

\author{Saroj Adhikari}
\affiliation{Department of Physics, University of Michigan, 
450 Church St, Ann Arbor, MI 48109-1040, \\
  Leinweber Center for Theoretical Physics, University of Michigan, 
  450 Church St, Ann Arbor, MI 48109-1040}

\author{Dragan Huterer}
\affiliation{Department of Physics, University of Michigan, 
450 Church St, Ann Arbor, MI 48109-1040, \\
  Leinweber Center for Theoretical Physics, University of Michigan, 
  450 Church St, Ann Arbor, MI 48109-1040}

\date{\today}

\begin{abstract}
Several unexpected features are observed at large angular scales in the cosmic microwave background (CMB) temperature anisotropy measurements by both WMAP and Planck. These include the lack of both variance and correlation, alignment of the lowest multipole moments with one another, hemispherical power asymmetry, and an odd-to-even parity excess.  In this work, we study the statistics of eight representative large-angle CMB features in order to evaluate their covariance in the standard $\Lambda$CDM model. We do so using two sets of simulated CMB temperature maps; an ensemble of 100,000 simple Gaussian simulations, and 1000 Full Focal Plane (FFP) simulations  provided by the Planck collaboration. In measuring feature probabilities, we pay particular attention to analysis choices, making sure that we can reproduce previous  results in the literature, and explain differences where appropriate. The covariance structure we find is consistent with expectations given that many of the features studied are functions of the angular power spectrum. Notably, we find significant differences in the covariance entries associated with the quadrupole-octopole alignments derived from the Gaussian and FFP simulations. We additionally perform a principal component analysis  to quantitatively gauge what combinations of features capture the most information about how simulation measurements vary, and to provide an alternative assessment of the ways in which the real sky is anomalous. The first four principal components explain about 90\% of the simulations' variance, with the first two roughly quantifying the lack of large-angle correlations, and the next two quantifying the phase-dependent anomalies (multipole alignments and power asymmetry). Though the results of this analysis are fairly unsurprising, its comprehensive  approach serves to tie together a number of previous results, and  will therefore provide context for future studies of large-angle anomalies.
\end{abstract}
\maketitle
\section{Introduction}\label{sec:intro}

The spectacular maps of the cosmic microwave background (CMB) anisotropy  that have been made over the past few decades have
revolutionized our understanding of the universe, and rejuvenated efforts to
test fundamental processes in the early and late universe. The CMB maps are
overall in a very good agreement with the six-parameter spatially flat $\Lambda$CDM model
specified by the energy densities of dark matter and baryons, the amplitude
and spectral index of primordial scalar fluctuations, the reionization optical
depth, and the expansion rate (Hubble constant) \cite{Ade:2015xua}. Shortly
after the WMAP experiment's data were released, however, several surprising
coincidences were noticed at large angular scales. In particular, the WMAP
maps of temperature anisotropies exhibit low variance, a lack of correlation
on the largest angular scales, alignment between various low multipole
moments \cite{deOliveira-Costa:2003utu}, alignment between those low multipole
moments and the motion and geometry of the Solar System \cite{Copi:2006tu}, a
hemispherical power asymmetry \cite{Eriksen:2003db}, a preference for odd
parity modes \cite{Land:2005jq}, and an unexpectedly large cold spot in the 
Southern hemisphere \cite{Vielva:2003et}. Planck data \cite{Ade:2015hxq} 
largely confirmed the presence of these features. For a review of the CMB 
anomalies, see Ref.~\cite{Schwarz:2015cma}.

While these
anomalies have remained an active area of study over the years, it is
difficult to draw firm conclusions from the study of features of the CMB at
very large angles, mainly due to the significant cosmic variance at those
scales. Moreover, the \textit{a posteriori} nature of their observation, as
well as the generally good fit of data to the standard cosmological model,
means that large-angle anomalies do not in themselves provide compelling
evidence for beyond-$\Lambda$CDM physics~\cite{Bennett:2010jb}.

Rather, in a time  where nearly all cosmological
observations have been in remarkable agreement with the predictions of
$\Lambda$CDM, the statistically unlikely large-angle features have attracted
attention because of the tantalizing possibility that one or some of them
might have cosmological origins \cite{Contaldi:2003zv, Gordon:2005ai,Bunn:2008zd,Adhikari2016}.  If that were the case, due to e.g.\ an
isotropy-breaking mechanism in the early universe, the feature in question
could provide insight into the physics of inflation. It is important, however, 
to additionally consider other explanations for anomalous
large-scale CMB features: they could be artifacts of instrumental or
astrophysical systematics or they could simply be unlikely fluctuations in
the standard isotropic model. Much of the study of large-angle anomalies has
thus been focused on disentangling these three logical possibilities: whether
large-angle CMB anomalies are cosmological,  are due to systematics, or
are statistical flukes. Better understanding of the anomalies  will
be driven in the future by observations of new quantities on very large spatial scales, such
as CMB polarization \cite{Dvorkin:2007jp, Copi:2013zja,
  Yoho:2015bla,ODwyer:2016xov, Contreras:2017zjv} and lensing
\cite{Yoho:2013tta}, as well as large-scale structure \cite{Gibelyou:2010qe}. Whether or not
new insights about the early universe become readily apparent, studying
large-angle anomalies has and will thus continue to provide  an opportunity
to build a deeper understanding of our measurements of the large-angle CMB.

One largely unexplored question is how the observed anomalies are related to
one another: If we observe one unlikely feature, does that make us less
surprised to find another? Roughly speaking,
(positively) correlated anomalies imply a smaller overall joint significance
than if they are uncorrelated. Full understanding of the anomalies thus
enables an accurate accounting of the likelihood for the joint observation of
unexpected features. Since the {\it a posteriori} nature of the choice of 
anomaly statistics remains, we would certainly not advocate for using a measure 
of the combined tension of all observed anomalies as a  test of $\Lambda$CDM. 
However, for studies focusing on pairs or groups of large-angle features as 
tests for phenomenological models of inflation, knowing their joint probability 
distribution in $\Lambda$CDM will prevent double-counting (or, if
the anomalies are anticorrelated, undercounting) of the anomalies'
significance.
Furthermore, the knowing the covariance of large-angle features associated with anomalies allows us to quantitatively separate
anomaly ``atoms'', i.e. a set of independent features out of which drive the
anomalous observations~\cite{Schwarz:2015cma}.

Previous work on the covariance of CMB temperature anisotropy anomalies has
mostly been limited to studying pairs of anomalies. For example in
Ref.~\cite{Sarkar:2010yj}, the authors show that missing power at large scales
quantified by $S_{1/2}$ and the quadrupole-octopole alignment are not
correlated in $\Lambda$CDM (such a conclusion was also reached, albeit for the
full-sky-only analysis, by Ref.~\cite{Rakic:2007ve}). The lack of correlation between
hemispherical power asymmetry and the quadrupole-octopole alignment in
$\Lambda$CDM are demonstrated in Ref.~\cite{Polastri:2015rda}. In
Ref.~\cite{Kim:2010st}, the authors indicate a possible connection between the
lack of power and the odd-multipole preference anomaly. In particular, they
claim --- based on an analytical argument --- that the odd-multipole
preference can be a phenomenological cause of the lack of large-angle
correlation. Ref.~\cite{Gruppuso:2017nap} also studies this relationship by
examining how lack of power manifests separately in even and odd multipoles, finding that the excess odd power is contributed mainly by regions close to the galactic plane.
Additionally, in Ref.~\cite{Knight:2017kvk} the authors find a
correlation between the low large-angle power and the low value of the CMB
quadrupole.  Ref.~\cite{Copi:2008hw} explores the relationship between low
power at large angles and the amplitude of the CMB quadrupole and octopole,
while Ref.~\cite{Hajian:2007pi}  studies its relationship with the
quadrupole and octopole phases. Here we aim to take a more global view by
studying the relationship between all of these features simultaneously.

In this work, we will use ensembles of simulated CMB temperature maps to
empirically characterize, in the context of $\Lambda$CDM, the covariance between a collection of features associated with 
commonly-studied large-angle anomalies. Our analysis proceeds in three general
steps. First, we will measure the quantities associated with those features
and confirm that the comparison between our measurements of the real CMB sky
and simulations reproduce previous findings. Next, we will study the distribution of the simulation
ensembles in the space defined by the ``anomaly feature'' quantities to find
their covariances. In doing so, we will investigate the impact of foregrounds
and survey properties  by comparing the results obtained from simple Gaussian
simulations of the CMB temperature map to the more realistic Planck full focal
plane simulations~\cite{Ade:2015via}. Finally, we will use the measured
feature covariances perform a principal component analysis in order to further
characterize the ways in which large-angle CMB map properties are expected to
vary in $\Lambda$CDM, and in which the observed CMB sky is unusual. We
emphasize that the goal of this analysis is to gain a deeper understanding of the predictions of $\Lambda$CDM, rather
than to do any explicit model-comparison.

The rest of the paper is organized as follows. In \sect{sec:methods} we
introduce our methods in detail, including the description of maps and masks
adopted, of our simulation ensembles and of our power spectrum
measurements. In \sect{sec:anomalydef}, we outline the statistical description
of the eight large-scale features that we study in this work and report
the statistical significance of the features with respect to our simulation
ensembles. The main results of our work --- the measurement of covariances
between the anomalies --- is presented in \sect{sec:results}, along with
discussion of a PCA analysis. We summarize and conclude in \sect{sec:concl}.


\section{Methods}\label{sec:methods}

We begin by introducing basic
terminology and notation for describing CMB anisotropies.
Temperature fluctuations can be expanded in a spherical harmonic series as follows
\begin{equation}
T(\nhat) = \sum_{\ell}\sum_{m=-\ell}^{\ell} a_{\ell m} Y_{\ell m}(\nhat),
\end{equation}
where $\nhat$ is the direction on the sky and the complex coefficients $\alm$ contain all information about the temperature field. For statistically isotropic fluctuations, the expectation of the two-point
correlation between coefficients $\alm$ drastically simplifies and only depends on $\ell$:
\begin{equation}
\langle a_{\ell m} a_{\ell' m'}^* \rangle = \delta_{\ell\ell'} \delta_{m m'} C_\ell.
\label{eq:Cl}
\end{equation}
If the fluctuations are statistically isotropic and Gaussian, the angular power spectrum,  $C_\ell$, contains all statistical information about the temperature field.

It is useful to additionally define the real-space angular correlation function  as
\begin{align}\label{eq:cthetadef}
  C(\theta) &= \langle T(\nhat_1)T(\nhat_2)\rangle \\
  &= \frac{1}{4\pi}\sum_{\ell}(2\ell+1)\Cl P_{\ell}(\cos{\theta})
\end{align}
where $\nhat_1\cdot\nhat_2 = \cos{\theta}$ and $P_{\ell}(x)$ is a Legendre polynomial of order $\ell$.

\subsection{Maps and masks}
In the course of this analysis we use several data products from the Planck 2015 data release, provided on the Planck Legacy Archive\footnote{\tt pla.esac.esa.int}. For transparency and reproducibility, when relevant we identify the names of specific files used  in footnotes. 

Though the primary product of this paper will be a study of the covariance
between anomalies as measured from simulation ensembles, we also use Planck
data to quantify the values of the anomaly statistics described above for the
real, observed CMB sky. Our purpose in doing this will be twofold. First, it
will allow us to compare our assessment of the anomalousness  of features of the
observed CMB sky  against the probabilities reported in the
literature. Additionally, measuring the same statistics from
the real Planck data as from our simulations will allow us to place the real
CMB sky in the multidimensional feature space examined in
\sect{sec:anomalydef} and~\ref{sec:results}.

For map-based statistics we use the SMICA~\cite{Adam:2015tpy} map\footnote{\tt COM\_CMB\_IQU-smica\_1024\_R2.02\_full.fits} from the
2015 Planck data release.
Though the Commander map should more properly be used for the analysis of
very large scale features, past
studies \cite{Ade:2013nlj,Copi:2013cya,Ade:2015hxq} have found that the
significance of the various anomalies does not depend strongly on which
component separation method is used. Therefore, using the SMICA map should be sufficient for our purposes.

Because we care only about large-angle features, we work with maps that 
are at a resolution of $\nside=64$, smoothed with a Gaussian beam of 
160 arcmin. We  downgrade the SMICA CMB temperature map, which is 
provided at $\nside=1024$, following the prescription described 
in Ref.~\cite{Ade:2013nlj}.  We do this by  first extracting its spherical 
components $\alm$ using the {\tt healpy}\footnote{{\tt healpy} is the Python implementation of {\healpix}, described at {\tt 
http://healpix.sourceforge.net}.}~\cite{Gorski:2004by}   function {\tt map2alm}. 
Then, again using {\tt healpy}, we get the harmonic space representation of the 
Gaussian beam $b_{\ell}$ and pixel window functions $p_{\ell}$ corresponding to 
the full width half maximum (FWHM) and pixel resolution, respectively, of both 
the input and output maps. By combining these together, we obtain the 
downgraded harmonic coefficients,
\begin{equation}\label{eq:downgradealm}
\alm^{\rm out} = \frac{b_{\ell}^{(\rm out)}p_{\ell}^{(\rm out)}}{b_{\ell}^{(\rm in)}p_{\ell}^{(\rm in)}}\alm^{\rm in}.   
\end{equation}
We then use the {\tt healpy}~function {\tt alm2map} to convert back to pixel space, obtaining the downgraded map. We refer to Table 1 in Ref.~\cite{Ade:2013nlj} for the appropriate beam FWHM values: 160 arcmin for $\nside=64$, and 10 arcmin for $\nside=1024$. (For other parts of this study we will also use the conversions 5 arcmin for $\nside=2048$ and 640 arcmin for $\nside=16$.)

When we use a mask, we adopt the UT78 common mask\footnote{{\tt
    COM\_Mask\_CMB-IQU-common-field-MaskInt\_2048\_R2.01.fits}, field 0.},
which is identified in Ref.~\cite{Ade:2015hxq} as the mask that should be used
for the analysis of Planck temperature maps. UT78 is the union of the masks for
Planck's four component-separation methods (SMICA, NILC, SEVEM, and
Commander). This mask is provided  as a map of zeros and ones at resolution  $\nside=2048$,
where zeros represent masked pixels and ones signify unmasked pixels. To
downgrade the mask to $\nside=64$, we follow the procedure described in
\eq{eq:downgradealm}, then threshold the resulting map so that all pixels
with a value $\leq 0.9$ are marked as masked. This reduces the mask from its
original $\fsky=0.78$ to 0.67. (When we use $\nside=16$ maps for one of the
anomalies studied below, the UT78 mask's sky coverage reduces further to $\fsky=0.58$.)

\subsection{Simulated ensembles}\label{sec:ensembles}

Our primary simulation ensemble will be a set of 100,000 noiseless Gaussian CMB 
temperature maps generated using the {\tt 
  synfast} function in {\tt healpy}. Gaussian temperature map realizations are drawn using the  Planck best-fit theory prediction for the power spectrum.\footnote{{\tt  COM\_PowerSpect\_CMB-base-plikHM-TT-lowTEB-minimum}\dots{ }\dots {\tt-theory\_R2.02.txt}}. The maps are produced at $\nside=64$ with FWHM=160 arcmin Gaussian smoothing, and with the {\tt pixwin} argument set to {\tt True}. These settings were chosen to make the simulated maps have properties consistent with the downgraded SMICA temperature maps described above. 
These straightforward-to-implement simulations, which we will henceforth refer to as the 
``synfast simulations,'' will allow us to obtain the statistics of fluctuations
associated with the CMB signal only.

In order to explore whether foreground and survey-related effects that are present in the Planck SMICA map influence the relationship between anomalies, we repeat our analysis on the publicly available ensemble of Planck full focal plane (FFP) simulations~\cite{Ade:2015via}. 
Specifically, we use the FFP8.1 CMB sky
and noise maps that have been processed using the SMICA component separation pipeline, which we add together before downgrading to $\nside=64$.
The FFP simulations include the physical effects of astrophysical foregrounds, gravitational lensing, Doppler modulation, and frequency-dependent Rayleigh scattering effects. They also model the Planck mission's scanning strategy, detector response, beam
shape, and  data reduction pipeline. Additionally, a small, frequency-dependent intensity quadrupole has been added to the FFP simulations to account for an uncorrected residual in the data from the dipole-induced Doppler quadrupole identified in Ref.~\cite{Kamionkowski:2002nd}. Note that because we use the
FFP8.1 rather than FFP8 simulations, we do not need to rescale the CMB
components of the simulations by the factor of 1.0134 that was applied in
Ref.~\cite{Ade:2015hxq}. 

\subsection{Power spectrum measurements}\label{sec:clmeas}

\begin{figure*}
\includegraphics[width=.45\linewidth]{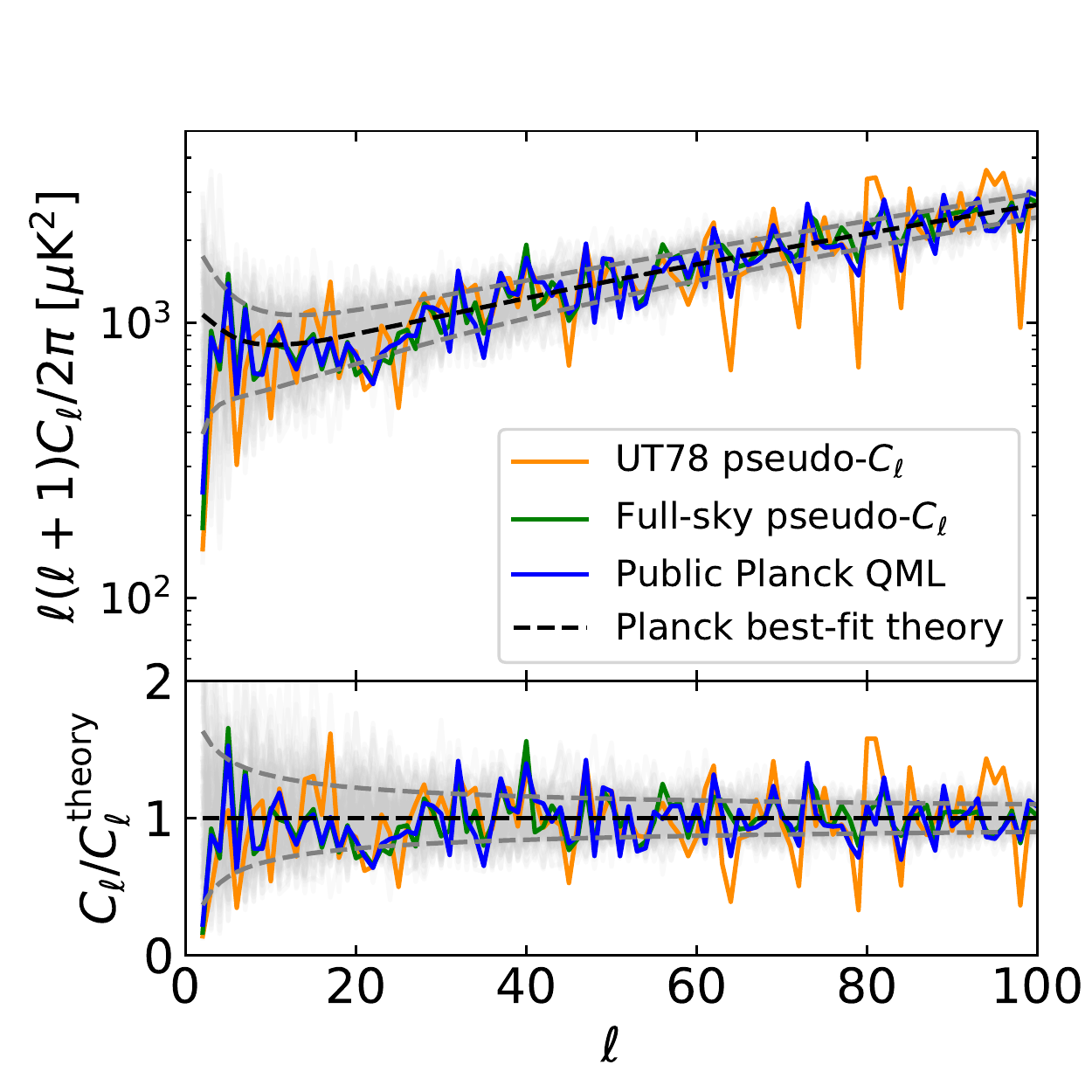}\includegraphics[width=.55\linewidth]{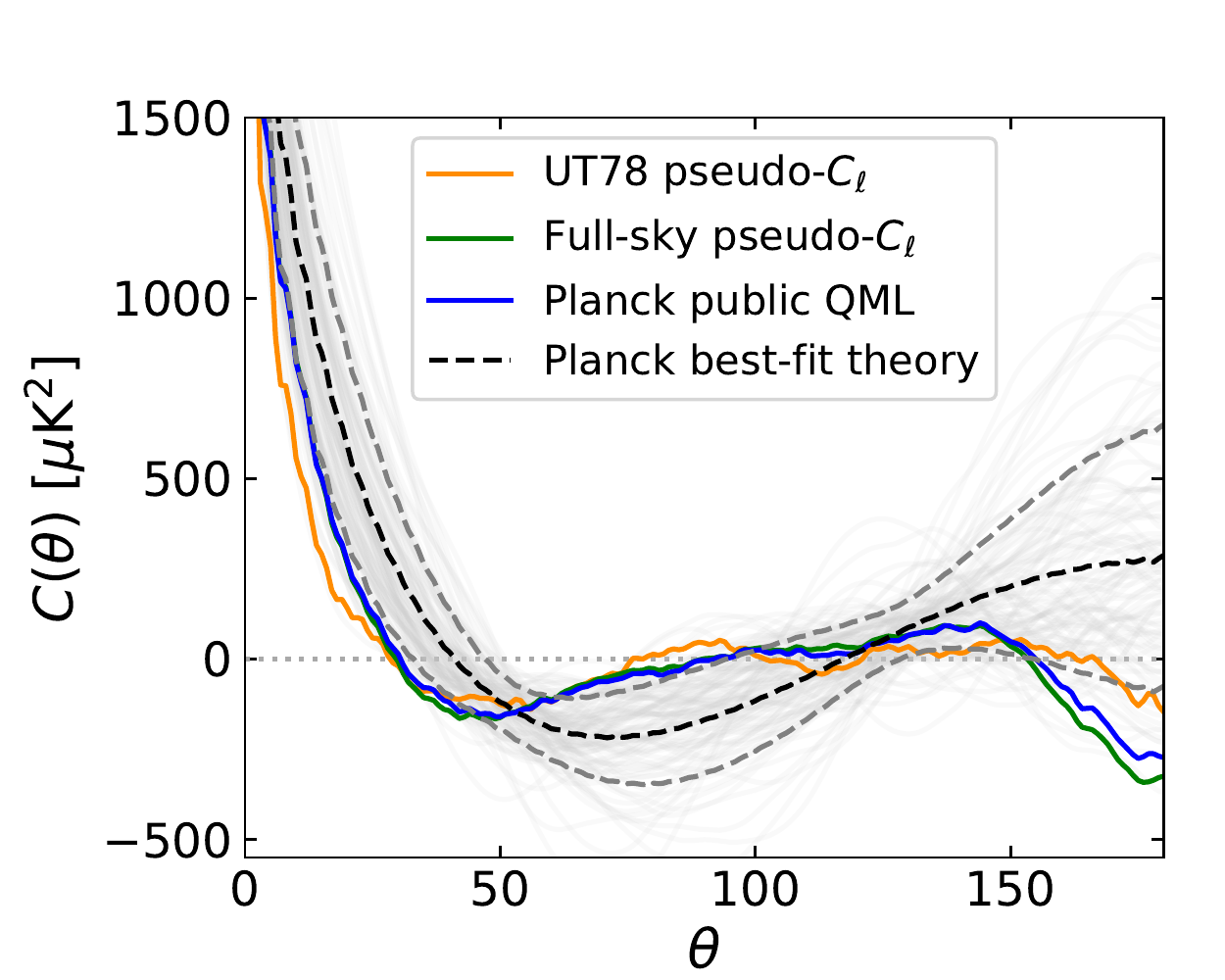}
\caption{Comparison of the different angular power spectrum measurements
  described in \sect{sec:clmeas} (left) and their corresponding angular correlation functions (right). The collection of grey lines behind them are from the full-sky $\Cl$ measurements of the first 100 synfast simulations. The black dotted line shows theoretical expectation, and the gray dotted lines show the  68\% confidence level cosmic-variance errors.}
\label{fig:clctcomp}
\end{figure*}

There are several methods that one can use to measure the angular power spectrum of a map of CMB temperature fluctuations. Using different methods generally will cause variations in the estimate for $\Cl$, and thus the choice of how to measure $\Cl$ can impact anomaly statistics.

 Our analysis mainly relies on pseudo-$\Cl$ estimates for the power spectrum. For full-sky measurements these will give unbiased estimates of the true power spectrum simply by averaging the observed spherical harmonics,
 \begin{equation}\label{eq:clsumdef}
\hat{C}_{\ell} = \frac{1}{2\ell+1}\sum_{m=-\ell}^{m=+\ell}|\alm|^2.
 \end{equation}
 If a mask is used to remove contaminated parts of the map, additional care must be taken, as is described in e.g. Ref.~\cite{Copi:2013cya}. 
 In practice we measure cut-sky pseudo-$\Cl$'s using  the {\tt polspice} algorithm\footnote{The polspice software can be found at {\tt http://www2.iap.fr/users/hivon/software/PolSpice/}. we run it using the settings {\tt subav=YES}, {\tt subdipole=YES}, {\tt apodizesigma=NO}, and {\tt pixelfile=NO}.}~\cite{Chon:2003gx}, which removes the monopole and dipole of the masked map, measures the angular correlation function $C(\theta)$ of the unmasked part of the sky, and then integrates to obtain $\Cl$,
 \begin{equation}\label{eq:clintdef}
   \Cl = 2\pi \int_{-1}^1\,C(\theta)\,P_{\ell}(\cos{\theta})\,d\cos{\theta}.
\end{equation}

In contrast to these cut-sky pseudo-$\Cl$'s, which only estimate the statistical properties of the
unmasked parts of the sky, quadratic maximum likelihood (QML)
methods can be used to estimate the statistical properties of the entire sky. 
QML power spectrum estimators are unbiased, and have a smaller variance than 
pseudo-$\Cl$ estimators~\cite{Molinari:2014wza}.
It will make the most sense for us to study the statistical properties of 
certain large-angle features in terms of the  observed SMICA map's QML power 
spectrum. For this, we do not implement our own QML power spectrum estimator,
but instead use the public QML spectrum provided by the Planck
team~\cite{Aghanim:2015xee}. The Planck QML power spectrum was obtained using a 
Blackwell-Rao estimator~\cite{Chu:2004zp} applied to the Commander 
component-separated
map (and mask) at $\nside=16$ for multipoles $\ell=2-29$, and the {\tt PliK}
likelihood applied to measured pseudo-$\Cl$'s for $\ell\geq 30$. The
low-$\ell$ power spectrum estimation uses the Commander mask, which has
$\fsky=0.94$ and therefore leaves available
  much more of the sky available for analysis than the UT78 common mask. The
high-$\ell$ power spectrum likelihood uses
galactic masks, described in  Appendix A of
Ref.~\cite{Aghanim:2015xee}, which  leave less available  sky than those used with
component-separated CMB maps.

To summarize, the three $\Cl$ measurement strategies we will examine are:
\begin{itemize}
\item {\bf Full-sky $\Cl$'s:} We computed them using {\tt polspice}
  based on a map with $\nside=64$, with the monopole and dipole subtracted. 
\item {\bf UT78 pseudo-$\Cl$'s:} We computed them using {\tt polspice} with
  the same settings as the full-sky case, using the $\nside=64$ version of the
  Planck UT78 common mask.
\item { \bf Planck public QML $\Cl$'s: } These estimates are provided for
  low\footnote{\tt COM\_PowerSpect\_CMB-TT-loL-full\_R2.02.txt} and
  high\footnote{\tt COM\_PowerSpect\_CMB-TT-hiL-full\_R2.02.txt}-$\ell$ on the
  Planck Legacy archive.
\end{itemize}
In \fig{fig:clctcomp} we compare the angular power spectra derived from the
SMICA CMB temperature map using these three methods, as well as the corresponding angular correlation functions derived using \eq{eq:cthetadef}. We also show theoretical predictions using the Planck best-fit model, along with the 68\% confidence region for cosmic variance.

Our $\nside=64$ resolution implies that we can study multipoles up to
$\ell_{\rm max} = 3\nside -1 = 191$ (though for practical purposes,
pixelization effects become apparent for pseudo-$\Cl$ measurements at
$\ell\sim 150$ for full-sky measurements and $\ell\sim 100 $ for
cut-sky). Since we will be focusing on scales $\ell<100$, this choice of
$\nside$ is sufficiently high.

\section{Features Studied}\label{sec:anomalydef}

Here we study eight characteristics of the large-angle CMB temperature
maps. These features, which are summarized in
Table~\ref{table:anomalies_list}, include some of the most prominently
discussed CMB anomalies.  This set of features is not intended to be
comprehensive\footnote{For example, we do not include
  the statistics from Ref.~\cite{Aluri:2017mhj} which quantify
  alignments between different combinations of even and odd parity
  multipoles. We also do not include the ``cold spot''
    \cite{Vielva:2010ng} which, being a localized feature at smaller scales,
    does not naturally belong to the set of large-angle features studied
    here.}  Rather, our intention is to focus on a representative sample that
will allow us to develop an understanding of the large-angle CMB's statistical
properties in $\Lambda$CDM.

Broadly, we classify features based on
whether they depend entirely on information in the (isotropic) two-point 
statistics, or whether they require map or $\alm$-based information. We adopt 
this
classification to aid in our interpretation of their covariances: because the
two-point function anomalies are all functions of the same angular power
spectrum, their respective definitions directly imply some \textit{a priori}
expectations for their covariances. The same is not necessarily true for the
isotropy-breaking anomalies, due to the stochastic nature of the $\alm$ coefficients.

As we introduce each feature, we will 
define the quantity that we use to measure it, briefly introduce relevant 
findings from
previous studies, and discuss how those findings compare to our
measurements. We will mainly perform these comparisons against
Ref.~\cite{Ade:2015hxq}, the Planck 2015 paper on the isotropy and statistics
of the CMB, which we will henceforth refer to as \PIS. Unless otherwise noted, the anomaly measurements in that work were
done using a QML $\Cl$ estimator on UT78 cut-sky maps evaluated at
$\nside=64$. Their measurements of the real sky were done on the SMICA
temperature map, and they evaluated statistics based on the FFP8
simulations. Note that because we use a different ensemble of simulations (FFP8.1 rather than FFP8), as
well as a different power spectrum measurement technique, we expect our
findings for anomaly statistics to be similar to the \PIS{} results, but
not necessarily to  exactly match them.

Following \PIS, we will quantify how unusual (or not) the SMICA temperature
map appears compared to simulations using $p$-values, defined to be equal to
the fraction of simulations that return more extreme
values than the real sky.  As part of each feature description, we will note
whether and how measurement choices (between Planck QML vs. pseudo-$\Cl$,
cut-sky vs. full-sky) affect those statistics, and will take care to identify
which of those choices are used in our anomaly covariance studies presented in
the next Section (\ref{sec:results}).  These single-feature results are
summarized in \fig{fig:1dhists}, which is described in more detail in
\sect{sec:1dstats}.

\begin{table*}
	\centering
	\begin{tabular}{l|c|l|c|c}
		\hline
		\textbf{Depends on} & \textbf{Quantity} & \textbf{Description} & 
		\textbf{Multipoles} & \textbf{Section} \\ 
		\hline\hline
		Two-point functions only & $\Shalf$ & Amount of angular power 
		at $\theta>60^\circ$ & $2-100$ & \ref{sec:S12}\\
		& $\quadrupole$ & Quadrupole amplitude & 2 &\ref{sec:quadrupole}\\
                & $\octopole$ & Octopole amplitude & 3 &\ref{sec:octopole}\\
		& $\svar$ & Variance of temperature fluctuations at 
		$N_{\rm side}=16$ & $2-47$ & \ref{sec:lowvar}\\	
		& $\Rparity$ & Ratio of power between even and odd multipoles & $2-27$ &\ref{sec:parity} \\
                & $\ctpi$ & Angular correlation at $\theta=180^{\circ}$ & $2-191$ &\ref{sec:ct180} \\ 
		\hline	
		Phases of $\alm$ & $\SQO$  & Quadrupole-octopole 
		alignment & $2,3$ &  \ref{sec:SQO}\\
		& $\ALV$ & Hemispherical power asymmetry & $2-191$ &\ref{sec:ALV} \\
		\hline \hline
	\end{tabular}
	\caption{Summary of quantities studied in this work.}
	\label{table:anomalies_list}
\end{table*}
\subsection{Features depending on two-point functions only} \label{sec:Clanomalies}
We first study the six features that are fundamentally a function of the angular
clustering power.
\subsubsection{$\Shalf$: Large-angle power}\label{sec:S12}

First, we measure power in large angular scales of the temperature map using the $\Shalf$ statistic, defined as the integral of the
square of the angular correlation function $C(\theta)$ over angles between
$60^\circ$ and $180^\circ$~\cite{Spergel:2003cb}
\begin{equation}
  S_{1/2}  =  \int_{-1}^{1/2} \left[ C(\theta)\right]^2 d(\cos \theta).
\label{eq:S12}
\end{equation}
It measures the deviation of $C(\theta)$ from zero at angles greater than
$60^\circ$. The inclusion of this statistic is motivated by the lack of power at large angular scales $\theta \gtrsim 60^\circ$ first
observed by COBE \cite{Hinshaw:1996ut}, and later confirmed by WMAP
\cite{Spergel:2003cb} and Planck~\cite{Ade:2015hxq,Copi:2013cya}.
Though there are several ways of quantifying this lack of large-angle correlation, we adopt $\Shalf$ because it is the most commonly studied in the literature.

In practice, to measure $\Shalf$ for a temperature map, we first measure the angular power spectrum and then
calculate it in harmonic space via~\cite{Copi:2008hw}
\begin{equation}
S_{1/2} = \frac{1}{(4\pi)^2} \sum_{\ell, \ell'}(2\ell+1)(2\ell'+1) 
\mathcal{C}_\ell I_{\ell, \ell'}\left (\frac{1}{2}\right ) \mathcal{C}_{\ell'}.
\end{equation}
Here the matrix $I_{\ell, \ell'}$ is defined as
\begin{equation}
I_{\ell,\ell'}(x) = \int_{-1}^x P_{\ell}(x') P_{\ell'}(x') dx',  
\end{equation}
but is in practice computed using the recursion relation in Appendix A
of Ref.~\cite{Copi:2008hw}. We sum over values $\ell=2-100$.

When analyzing the SMICA map at $\nside=64$ with the UT78 mask, \PIS{} reports a low value for $\Shalf$ with a lower tail
probability\footnote{Value from Table 13 of Ref.~\cite{Ade:2015hxq}} of
$p=0.4\%$. That means they find that only 0.4\% of simulations have a lower value of
$\Shalf$ than the SMICA map. Our cut-sky $\Shalf$  measurements give similar
probabilities: $p=0.7\%$ for the fiducial synfast simulations and $p=0.5\%$
for the FFP simulations.\footnote{These $p$-values are weakly sensitive to whether the $\Cl$'s are corrected for resolution according to \eq{eq:clbeamcorr}: with that correction, the $p$-values for $\Shalf$ go down to $0.6\%$ for the synfast simulations and to $0.4\%$ for the FFP simulations. We opt not to make that correction when computing $\Shalf$ and  $C(\pi)$ because doing so introduces significant noise contributions at high multipoles and makes the sums involved overly sensitive to our choice of $\ellmax$. } 

The  lower-tail probability of the observed sky's $\Shalf$ value depends
dramatically on the method used to measure the angular power spectrum, increasing to $8\%$ for full-sky $\Cl$'s and to $6\%$ for Planck public QML $\Cl$'s (which, recall, effectively reconstruct the full-sky anisotropy field). This is consistent with results from previous studies~\cite{Copi:2008hw,Gruppuso:2013dba,Copi:2013cya} which have shown that the
relatively small amount of (nonzero) correlations on the full sky are
dominated by contributions from pixels close to the galactic mask. 

\subsubsection{$\quadrupole$: Quadrupole amplitude}\label{sec:quadrupole}

We additionally study $C_2$, the quadrupole of temperature fluctuations, which was first found to be low 
in COBE \cite{Bennett:1996ce} data, and later in WMAP
\cite{Bennett:2003bz,deOliveira-Costa:2003utu} and Planck
\cite{Aghanim:2015xee}. Analyses have shown that the lowness of the quadrupole is not particularly
significant \cite{ODwyer:2004vgx,Chu:2004zp,Slosar:2004xj}, so its value or lower-tail probabilities are not generally reported explicitly in the literature. Given this, we do not directly compare our measurement of $C_2$ to previous results, but do include it as one of our statistics in order to study its covariance with the
low angular power at large angles and other features.

Our one-dimensional study of $\quadrupole$'s statistics reflect the findings
in the literature. Our fiducial choice for the quadrupole  is
to adopt the Planck QML $C_2$ to represent the observed value, while for the
simulations we use the quadrupole from the full-sky $\Cl$'s.
In order to make the simulation measurements more directly  comparable to the QML power spectrum, we apply a correction for the $\nside=64$ maps' beam and pixel window functions via
\begin{equation}\label{eq:clbeamcorr}
  \Cl = (b_{\ell}^{(64)}p_{\ell}^{(64)})^{-2}\Cl^{\rm polspice}.
\end{equation}
Here,  $b_{\ell}^{(64)}$ and $p_{\ell}^{(64)}$ are the harmonic components of the beam and pixel window functions for the $\nside=64$ input map.

We find that $\quadrupole$ has a lower-tail probability of $5\%$ using the synfast simulations, and $6\%$ compared to the FFP simulations. Pseudo-$\Cl$ measurements of the SMICA map give slightly lower probabilities, with $p\sim 2\%$ for either full- or cut-sky measurements.

\subsubsection{$\octopole$: Octopole amplitude}\label{sec:octopole}

Though the value of the observed CMB temperature octopole amplitude is not anomalous (see e.g.\ \cite{Slosar:2004xj}),
we also include it in our study  because its behavior in relation to other features has the potential to be interesting. For example, 
Ref.~\cite{Hajian:2007pi} points out that contributions from the quadrupole
and octopole seem to be canceling the power from the rest of the sky, and that
a measure of large-angle power becomes less anomalous
when their contributions to the correlation function are
removed. Additionally, Ref.~\cite{Copi:2008hw} finds that the relationship between  several of the
lowest multipoles, certainly more than the just the quadrupole,
is responsible for the low observed $\Shalf$. Given this, we include $\octopole$ in our analysis because the relationship between 
$\octopole$, $\quadrupole$ and $\Shalf$ may reveal some interesting structure.

We perform our fiducial measurement of $\octopole$ in the same way as for
$\quadrupole$: we use the Planck QML $\Cl$'s for the observed temperature map,
and the beam-and-pixelization-corrected [according to \eq{eq:clbeamcorr}] full-sky  $\Cl$ measurements from simulations. Compared to the synfast and FFP simulations, the $p$-values for both the QML and full-sky SMICA measurements are $47-49\%$, while the cut-sky octopole is lower, with $p\sim 15\%$.

\subsubsection{$\svar$: Variance  at $\nside=16$}\label{sec:lowvar}

We study another indicator of large-angle power via $\svar$, the variance of unmasked pixels of a low resolution, $\nside=16$ temperature map. The variance of CMB temperature maps, especially at low spatial resolution,
has  been observed to be anomalously low in analyses of both
WMAP~\cite{Cruz:2010ud,Monteserin:2007fv} and Planck~\cite{Ade:2015hxq}
data. Planck measured the variance of unmasked pixel values with various resolutions, and found that the lowest investigated value of $N_{\rm side}=16$, with a $p$-value\footnote{This value is taken from Table 12 of
  Ref.~\cite{Ade:2015hxq}.} of 0.5\%, produced the most anomalously low variance. They found that the variance tends to
become lower as the mask is extended to cover more of the
sky, and that the statistical significance of its lowness persists when different foreground
subtraction methods are applied.

To measure $\svar$ for a given CMB temperature map, we first downgrade the map from
$\nside=64$ to $\nside=16$. We also downgrade the UT78 mask, but
go directly from the original $\nside=2048$ resolution to $\nside=16$ in order
to make the resulting sky fraction consistent with that used in the Planck
study. We then simply compute the variance of all unmasked pixels.

Though we measure $\svar$ through a pixel-based method, given an angular power spectrum $\Cl$ we can predict its expectation value for full-sky measurements via
\begin{equation}\label{eq:s16expect}
  \langle \svar \rangle(\Cl) = \frac{1}{4\pi}\sum_{\ell}(2\ell+1)\Cl\, (b_{\ell}^{(16)}p_{\ell}^{(16)})^2,
\end{equation}
where $b_{\ell}^{(16)}$ and $p_{\ell}^{(16)}$ are the beam and pixel window functions corresponding to $\nside=16$. This expression will allow us to compare our map-based measurements of $\svar$ to the predictions from the Planck best fit theory $\Cl$'s as well as the Planck public QML $\Cl$'s. 

It is worth noting that our method of measuring $\svar$ is different from that used to quantify map variance in the WMAP and Planck analyses. Those analyses use an  estimator~\cite{Cruz:2010ud} to isolate the cosmological contribution to the variance of a normalized version of the temperature map, in which each pixel value has been divided by its expected dispersion from both cosmological temperature
fluctuations and noise. Because of this, our reported numbers for $\svar$ will be much larger than the normalized variances reported in \PIS. Nonetheless, the statistical distribution of variances should be similar, to the extent that noise contributions to variance can be approximated as direction-independent.\footnote{In principle the noise dispersion can vary with position on the sky due to beam
effects and weights used to construct component separated maps, but
those effects are expected to be small~\cite{Monteserin:2007fv}.} 

For measurements of $\svar$ we would like to exclude pixels that may contain residual foregrounds, so we focus on its cut-sky value for both from the SMICA map and simulations. We find the SMICA $\svar$ to be low compared to simulations, with single-tail probability of $p=0.8\%$ and $0.5\%$ for the {\tt synfast} and FFP simulations, respectively. Thus, our cut-sky FFP $p$-value exactly matches that in \PIS. The $\svar$ expectation value from the Planck public QML $\Cl$'s and our our full-sky $\Cl$ measurements are very similar, with $p=20\%$ when compared to either simulation ensemble.

\subsubsection{$\Rparity$: Parity asymmetry at low $\ell$}\label{sec:parity}

We use the statistic $\Rparity$ to quantify large-angle parity asymmetry of the CMB temperature map. It has been noted that, at low $\ell$, the CMB maps have more power in odd
multipoles than even. This was observed in the WMAP 3, 5, and 7 year data~\cite{Kim:2010gd,Kim:2010gf,Grupposo2011} as well as in Planck \cite{Ade:2015hxq}.  We
quantify this asymmetry using the same estimator as \PIS,
\begin{equation}\label{eq:Rlmax}
R_{\ellmax} = \frac{D_+(\ell_{\rm max})}{D_-(\ell_{\rm max})}
\end{equation}
where
\begin{equation}
D_{+,-}= \frac{1}{\ell_{\rm tot}^{+,-}} \sum_{\ell=2, \ell_{\rm 
    max}}^{+,-} \frac{\ell(\ell+1)}{2\pi} C_\ell,
\label{eq:D}
\end{equation}
and the plus and minus indicate sums over even (parity-symmetric) and odd (parity-antisymmetric) multipoles, respectively.
The $R_{\ellmax}$ statistic is therefore a ratio of the parity-even over 
parity-odd multipole band-powers. The factor of $\ell(\ell+1)/(2\pi)$ is used 
because the theoretical prediction for $\ell(\ell+1)(2\pi)^{-1}\Cl$ is 
approximately scale-independent out to multipoles of $\ell\lesssim 50$, and 
thus gives  $R_{\ellmax}\sim 1$ over that range.

Because $R_{\ellmax}$ is directly based on the power
spectrum, we will focus on its measurements  from the Planck QML
power spectrum, and compare them to full-sky $\Cl$ measurements in
simulations. As in the case of $\quadrupole$ and $\octopole$, we correct for the impact of
the simulations' resolution on the power spectrum using
\eq{eq:clbeamcorr}.

For our covariance studies, we will focus on the behavior for  $\ellmax=27$, as that multipole range gives the most anomalously low value of $R_{\ellmax}$ in the \PIS{} analysis\footnote{Value taken from text associated with Fig.~20 of Ref~\cite{Ade:2015hxq}. Though that text actually reports $\ell=28$ to give the lowest $R_{\ellmax}$ $p$-value, this is due to a typographical error, and we confirmed with that section's author that the minimum $p$-value is actually at $\ell=27$.~\cite{Molinari:2018emails}}, with a single-tail probability $p=0.2\%$ for the SMICA map.  
We find the SMICA map's $\Rparity$ to be notably less anomalous: measurements of the Planck QML power spectrum give single tail probabilities of $p=3\%$ and $2\%$ when compared to the {\tt synfast} and FFP simulations, respectively.

\begin{figure}
\centering
\includegraphics[width=\linewidth]{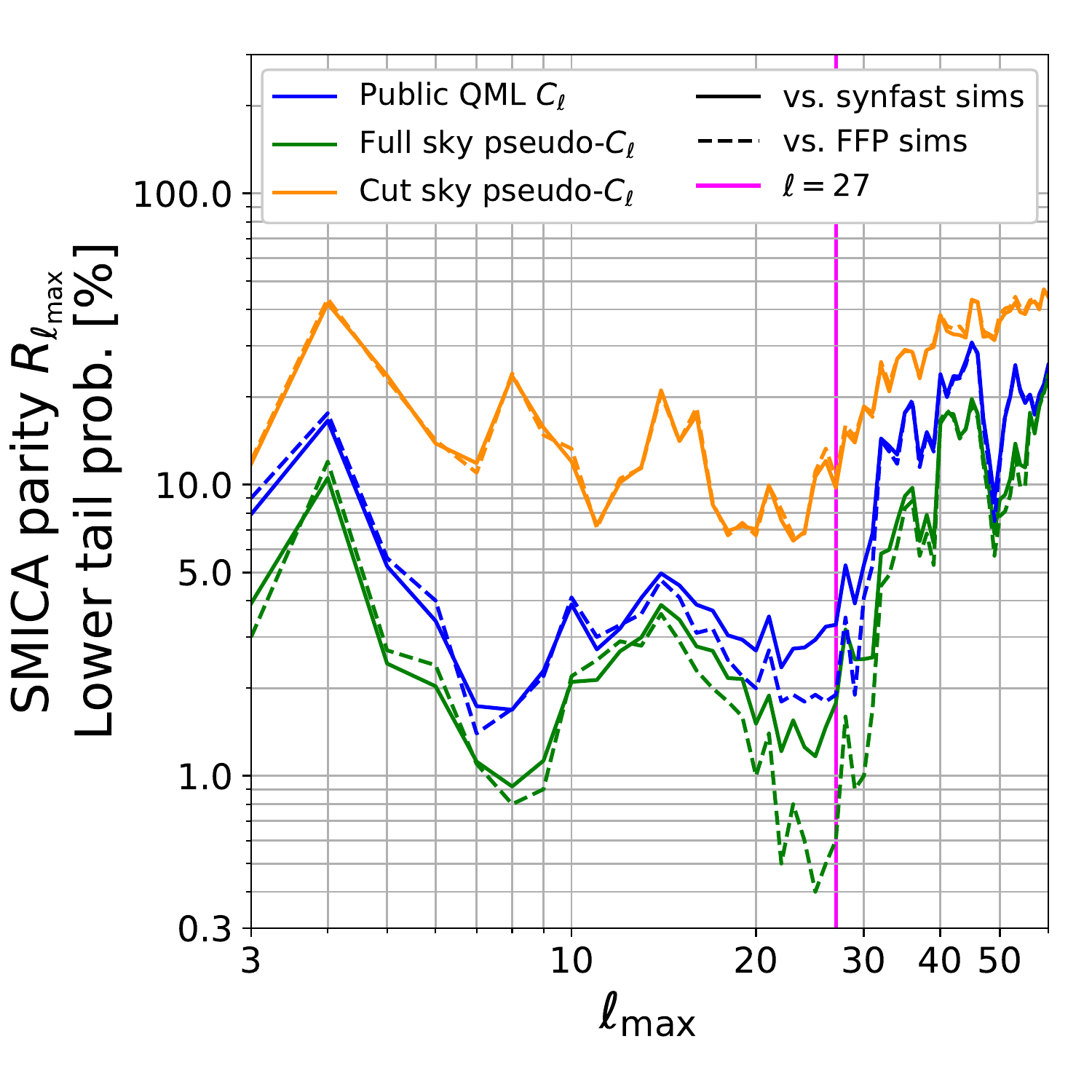}
\caption{Plot of the lower-tail probability $p$ for the parity asymmetry as a
  function of the largest multipole $\ellmax$ considered, for various SMICA
  power spectrum measurement and simulation ensemble combinations. Solid lines
  show the probabilities for the SMICA map assessed relative to the synfast simulations, and dashed lines show them relative to the
 FFP simulations. For the QML and full-sky $\Cl$ SMICA
  measurements, the simulations are measured using full-sky
  $\Cl$'s. For the cut-sky pseudo-$\Cl$ SMICA measurements, cut-sky
  pseudo-$\Cl$ simulation measurements are used. The vertical line
    denotes the $\ellmax$ value at which where \PIS{} found the most
    anomalous parity statistic; see text for details.}
\label{fig:parity_pvslmax}
\end{figure}

Given this $p$-value discrepancy, we investigated how $R_{\ellmax}$ depends on
$\ellmax$ and the power spectrum measurement technique. Results of this
investigation, shown in \fig{fig:parity_pvslmax}, reveal that the significance
of parity asymmetry heavily depends on the choice of mask and power spectrum
measurement method. The fact that measurements using the SMICA cut-sky pseudo-$\Cl$'s are less anomalous than those using the full sky SMICA $\Cl$'s or the QML $\Cl$'s (which attempt to reconstruct the full sky), reflects the findings of Ref.~\cite{Gruppuso:2017nap}: that including data from high galactic latitudes only  reduces the even-odd power asymmetry. 
The discrepency between our results and those of \PIS{}
can  be explained by two main differences~\cite{Molinari:2018emails}. The first and primary cause is that the \PIS{} measurements of $R_{\ellmax}$ for the real sky
are based on QML $\Cl$ measurements of the UT78 cut-sky map, degraded to $\nside=32$, which leaves $\fsky=0.64$ unmasked. This has a
smaller sky fraction than the maps used to produce the Planck public QML power
spectrum, which was obtained using the Commander $\chi^2$-based LM93 mask with $\fsky=0.93$~\cite{Adam:2015wua}. The other difference is that our simulation measurements are based on pseudo-$\Cl$ measurements of the full sky, while those in \PIS{} use QML simulation measurements of the cut sky, but we find that this has  little impact on the probability distribution for $R_{\ellmax}$. 
We confirmed that we were able
  to replicate the Planck results when using their UT78 QML $\Cl$ values,
  which are an intermediate data product of their analysis, but restrict all
  results presented in this paper to only publicly available data.

We obtain a
$p$-value closest to that reported for $\Rparity$ in \PIS{} using full-sky
$\Cl$ measurement of the SMICA map compared to FFP simulations
(0.6\%). However, in order to be consistent with our treatment of the other
purely power-spectrum based features, we will use the public Planck QML
$\Cl$'s compared to full-sky $\Cl$ simulation measurements as our
fiducial choices for measuring $\Rparity$.

\subsubsection{$\ctpi$: Two-point correlation at $\theta=180^\circ$}\label{sec:ct180}

We next consider the angular correlation function of CMB temperature evaluated
at $180^\circ$, which we will refer to as $C(\pi)$. We include it in the hope
that it will help clarify the relationship between other features. Our
motivation comes from the fact that $C(\theta)$, which is otherwise fairly
flat at large angles, drops to negative values at $\theta\simeq \pi$. This dip
has been observed in both WMAP and Planck data, and can be seen in the colored
curves on the right-hand side of \fig{fig:clctcomp}. By its definition we
expect the value $C(\pi)$ to be related to the missing large-angle
correlations statistic $S_{1/2}$, as well as to the $\Rparity$ measurement of
parity asymmetry. This can be seen by comparing the definition for
$R_{\ellmax}$ in \eq{eq:Rlmax} to
\begin{equation}\label{eq:ctpidef}
  C(\pi) = \sum_{\ell=2}^\infty (-1)^\ell\, \frac{2\ell+1}{4\pi} C_\ell.
\end{equation}
We will of course  investigate these correlations quantitatively further below.

We measure $\ctpi$ by using cut-sky pseudo-$\Cl$ measurements of the SMICA map
and simulations to obtain measured power spectra (as we do for $\Shalf$),
transforming to real space angular correlations via \eq{eq:ctpidef}. In
practice, we compute the sum over multipoles $\ell=2-100$. For the cut-sky
measurements, the SMICA map's $\ctpi$ is lower than for the majority of
simulations, but is not particularly anomalous, with $p\sim 11\%$ for both the
synfast and FFP simulation ensembles. The $p$-value goes down to $5-6\%$ if we
instead compare SMICA map measurements using the public Planck QML $\Cl$'s to the same set of simulation measurements,
and $3-4\%$ for full-sky $\Cl$'s.  As this measure is not commonly
studied in the literature, we do not compare this to any reported values.

\subsection{Features depending on $\alm$ phases} 
\label{sec:almanomalies}
We also consider two features that depend on the phases of CMB temperature 
harmonic coefficients $\alm$ --- i.e. which cannot be measured solely from the 
two-point correlations of the maps. These features are associated with reported 
anomalies that could indicate a possible departure from the assumption of 
statistical 
isotropy of CMB temperature fluctuations: the quadrupole-octopole alignment and 
the hemispherical power asymmetry. 

\subsubsection{$\SQO$: Quadrupole-octopole alignment}\label{sec:SQO}

The CMB temperature's quadrupole and octopole were first
observed to be planar and aligned in
Ref.~\cite{deOliveira-Costa:2003utu}. There are a number of possible ways to
denote the directionality, and thus alignment, of multipoles. To do so, we
will follow the approach presented in Ref.~\cite{Copi:2003kt} and make use of
Maxwell multipole vectors. Multipole vectors are a representation of a
function on a sphere; while they are at some level equivalent to spherical
harmonics in that role, their relationship to the $\alm$ is highly nonlinear
in a way that makes them particularly well suited to studying the
directionality of patterns on the sky.

For each multipole $\ell$, there are $\ell$ corresponding multipole vectors $\mathbf{v}^{(\ell, i)}$, where $i\in \{1, 2,\ldots, \ell\}$. Roughly speaking, the more planar the temperature fluctuations associated with a given multipole are, the more its associated multipole vectors will be confined to a plane, and the more the oriented-area vectors defined by their cross products, 
\begin{equation}
  \mathbf{w}^{(\ell, i, j)} \equiv \pm (\mathbf{v}^{(\ell, i)}\times \mathbf{v}^{(\ell, j)}),
\label{eq:oa_vectors}
\end{equation}
 will line up in a direction normal to that plane.  Moreover, planarity (as
 opposed to simply orientation along a direction) of the temperature multipole
 will cause the multipole vectors $\mathbf{v}^{(\ell, i)}$ and
 $\mathbf{v}^{(\ell, j)}$ to be at large angles relative to each other,
 enhancing the magnitude of $\mathbf{w}^{(\ell, i, j)}$. Thus, we can use the
 extent to which the object oriented vectors for two multipoles point in
 similar directions to measure how much the power from those $\ell$-modes are
 aligned.

The statistic $\SQO$ takes advantage of this property to 
quantify the quadrupole-octopole alignment. It is the normalized sum of the dot
products of the quadrupole oriented-area vector $\mathbf{w}^{(2, 1, 2)}$ with
the three octopole oriented-area vectors $\mathbf{w}^{(3, i, j)}$:
\cite{Schwarz:2004gk,Copi:2005ff}
\begin{equation}
  \SQO = \frac{1}{3} \sum_{\{i, j\}} |\mathbf{w}^{(2, 1, 2)}\cdot 
  \mathbf{w}^{(3, i, j)}|
\label{eq:S_stat}
\end{equation}
where $\{i, j\}$ can be $\{1, 2\}$, $\{2, 3\}$, or $\{3, 1\}$. Given this,
larger values of $\SQO$ indicate more alignment and planarity in the $\ell=2$
and $\ell=3$ modes of the temperature maps.

Because multipole vectors are defined in terms of $\alm$, this measurement can
only be done on full-sky maps. To measure $\SQO$ for a temperature map, we
first use the \healpix{} function {\tt map2alm} to measure its $\alm$, and
then use  the procedure\footnote{Calculations were performed using code {\tt
    mpd\_decomp.py} provided at {\tt
    http://www.phys.cwru.edu/projects/mpvectors/}.} described in Appendix A of
Ref.~\cite{Copi:2003kt} to extract the multipole vectors for $\ell=2-3$. We
then combine them via Eqs.~(\ref{eq:oa_vectors}) and (\ref{eq:S_stat}) to get $\SQO$.

We find that the $\SQO$ value measured from the SMICA map is larger than that from most
simulations, with a $p$-value (here, upper-tail probability) of $0.4\%$ when
compared with either our synfast or FFP simulation ensembles. This is
consistent with the results in Ref.~\cite{Copi:2013jna}. They found the SMICA map from the Planck 2013 data release \cite{Ade:2013sjv} had a larger $\SQO$, with a $p$-value\footnote{Value from Table~7 of Ref.~\cite{Copi:2013jna}}  of 0.54\% compared to an ensemble of $10^6$
simulations analogous to our synfast simulations, but which use 
constrained realizations to in-paint masked regions. The fact that our 
$p$-value is so similar to theirs indicates that simply measuring $\SQO$ on 
full-sky maps (as we do) rather than doing in-painting does not significantly 
affect the large-scale alignment behavior.

\subsubsection{$\ALV$: Hemispherical power asymmetry via local-variance dipole}\label{sec:ALV}

Finally, we include a measure of the level of asymmetry in temperature power
between two hemispheres of the sky. This is studied because one hemisphere of
the observed CMB sky has been noted to have more power than the other
\cite{Eriksen:2003db, Hansen:2004mj}, which can be modeled by a dipole
modulation of temperature fluctuations at large angular scales
\cite{Eriksen:2007pc, Hoftuft:2009rq}.  Following Refs.~\cite{Akrami:2014eta,
  Adhikari:2014mua}, we quantify hemispherical power asymmetry using a 
  local-variance map, which measures the size of temperature fluctuations within
disks of radius $\theta$ centered on each of its pixels. By measuring the
dipole of a local-variance map, we can quantify the direction and magnitude of
any hemispherical power asymmetry in a computationally inexpensive
way. Additionally, we can probe the scale dependence of the effect by varying
the angular size $\theta$ of the disks used to create the local-variance map.

More formally, if $\nhat_i$ is the location 
of the  $i$th pixel of the input temperature map $T(\nhat)$ from which the monopole and dipole of unmasked pixels have been removed, we can write the local-variance map $\sigma_{\theta}^2(\nhat)$ as 
  \begin{equation}
\sigma_{\theta}^2(\nhat) =\frac{1}{N[\mathcal{D}_{\theta}(\nhat)]}\sum_{i\in\mathcal{D}_{\theta}(\nhat)} \left[T(\nhat_i) - \bar{T}_{\theta}(\nhat)\right]^2 ,    
  \end{equation}
  where $\mathcal{D}_{\theta}(\nhat)$ is  the set of unmasked pixels 
within angle $\theta$ of direction $\nhat$, $N[\mathcal{D}_{\theta}(\nhat)]$ is the number of pixels in that set, and $\bar{T}_{\theta}(\nhat)$ is their average temperature. 

In practice, we measure the dipole of a dimensionless, weighted version of the
local-variance map, 
\begin{equation}
\tilde{\sigma}_\theta^2(\nhat) = \frac{w(\nhat)}{\bar{w}}\times\frac{\sigma_\theta^2(\nhat)- 
	\mu_{\theta}(\nhat)}{\mu_{\theta}(\nhat)}. 
\end{equation}
In this expression, $w(\nhat)$ is a dimensionless weight map (defined
below), $\bar{w}$ is its average over unmasked pixels, and $\mu_{\theta}(\nhat)$ is the mean 
computed by averaging the local-variance maps of an ensemble of simulated CMB 
temperature maps.

We choose the weight function $w(\nhat)$ to be
\begin{equation}
  w(\nhat) =\frac{1}{ \Var[\sigma_{\theta}^2(\nhat)]}\times \left\{\frac{1}{N_{\rm pix}}\sum_{i =1}^{N_{\rm pix}}{\Var}[\sigma_{\theta }^2(\nhat_i)] \right\}
\end{equation}
where we define, if $\alpha$ labels the simulation realization and $N_{\rm sim}$ is the number of simulations,
\begin{equation}
  {\Var}[\sigma_{\theta}^2(\nhat)] = \frac{1}{N_{\rm sim}}\sum_{\alpha=1}^{N_{\rm sim}} \left[\frac{\sigma^{2(\alpha)}_{\theta}(\nhat) - \mu_{\theta}(\nhat)}{\mu_{\theta}(\nhat)}\right]^2. 
\end{equation}
This inverse variance weighting suppresses the impact of noisy regions of the
input temperature map, as long as those noise contributions are modeled in the
simulations. In the limit that noise properties are direction-independent, the weight factor $w(\nhat)/\bar{w}$ will approach 1. 

In our work, we measure $\ALV$ from the UT78 cut-sky for both the SMICA and 
simulation maps. We fix the disk radius to be $\theta =8^\circ$, which is the 
scale previously found to produce the most anomalous local-variance dipole, 
and  compute the local-variance maps at a resolution of $\nside=16$. 
The amplitude of the dipole of a normalized
local-variance map, $\ALV$, is then obtained by using the Healpix function
$\texttt{remove\_dipole}$. Following \PIS, we include only disks for which at least 90\% of the input pixels are unmasked.

Our measurements return a local-variance dipole amplitude with $\ALV= 0.22$ when $\sigma_{\theta}^2(\nhat)$ is normalized using  the synfast simulations, and $\ALV=0.21$ using the FFP simulations. These values give an upper tail probability of $p=1\%$ when compared to either set of simulations. Both of these values are notably larger than the \PIS{} findings of $\ALV\sim 0.044$ for the SMICA map, with a $p$-value of 0.1\%.\footnote{Values taken from Fig.~27 ($\ALV$ value) and Table~20 ($p$-value) of Ref.~\cite{Ade:2015hxq}.}
In investigating this discrepancy, we found that the value of $\ALV$ is highly 
sensitive to the resolution of the input temperature maps, with lower 
resolution input maps tending to give larger dipole amplitudes. Because we 
compute local variances using our fiducial set of $\nside=64$ maps, while 
\PIS{} uses $\nside=2048$ input maps, we believe this resolution dependence 
explains the difference.

We also note that Doppler dipole modulation included in the FFP
simulations~\cite{Aghanim:2013suk} (but not the synfast simulations) will
generate a small power asymmetry which contributes to the local-variance
power asymmetry. However, that contribution is expected to be negligible for
the scales $\ell\lesssim 191$ that we are investigating
\cite{Quartin:2014yaa}.  The fact that the $p$-values from comparisons to the
synfast and FFP simulations are nearly identical is in line with that
expectation.

\subsection{Summary: Individual anomaly measurements}\label{sec:1dstats}

\begin{figure*}
\centering
\includegraphics[width=\textwidth]{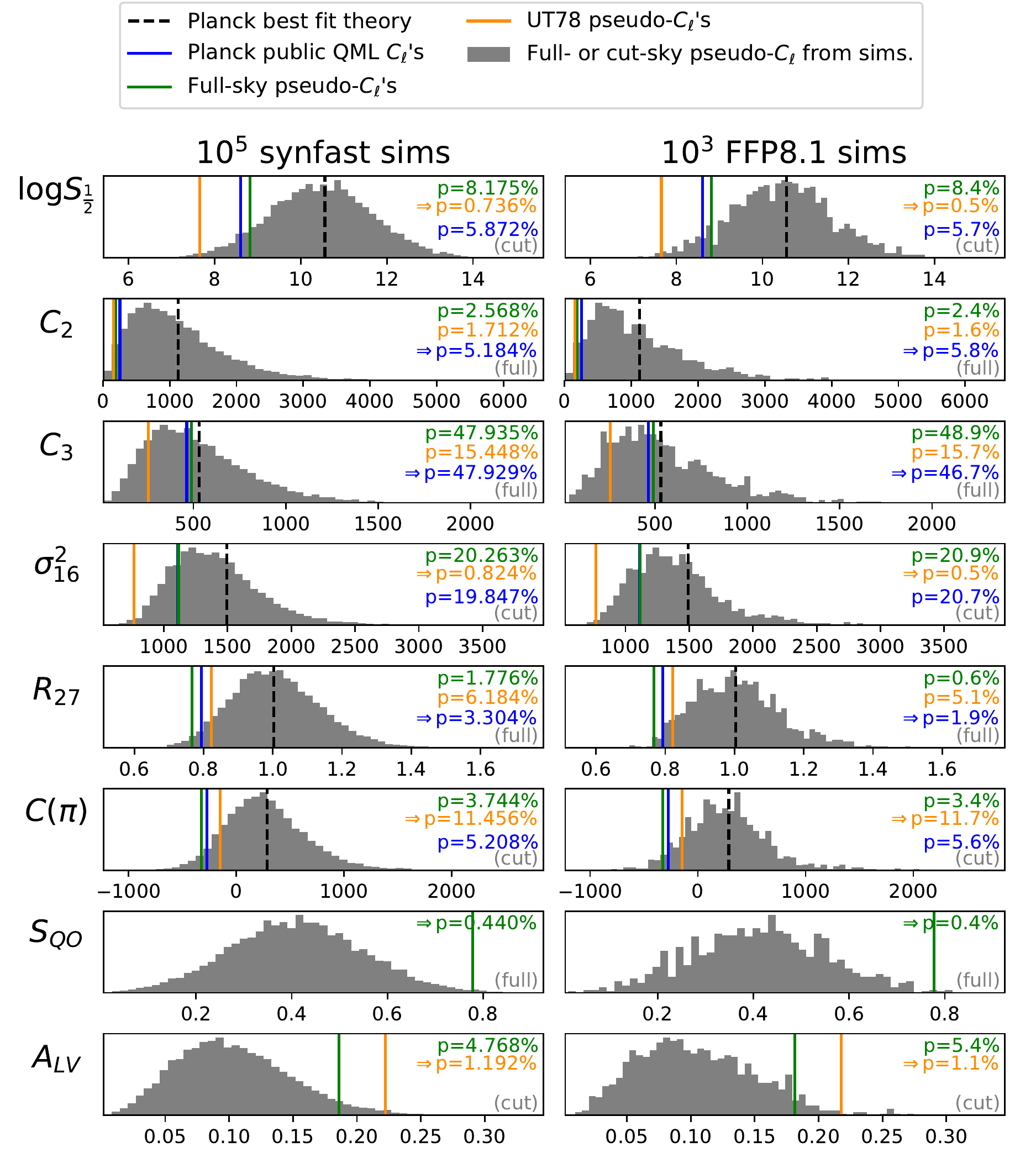}
\caption{Summary plot where each row shows the results for one of the features discussed in \sect{sec:anomalydef}.  Grey histograms show statistics for the features of the CMB temperature map measured from our 100,000 synfast simulations (left column), and from  1,000  FFP 8.1 simulations (right column). The grey text in the lower right of each panel denotes whether simulations measurements are based on the full-sky or the cut-sky, where cut-sky measurements use the UT78 mask. Vertical lines show values for the Planck SMICA map, as well as theory predictions from Planck's best-fit cosmology. The $p$-values displayed are single-tail probabilities showing the percentage of simulations that are more extreme that the  SMICA measurements. Arrows by the $p$-values indicate which measurement we think is most relevant for each feature.}
\label{fig:1dhists}
\end{figure*}

To summarize, we have defined quantities associated with eight properties of
the CMB temperature map which are either found to be statistically unlikely in
the observed sky or which are expected to shed light on the relationship
between statistically unlikely features. For each feature, we have described
our technique for measuring it from real and simulated CMB maps. By analyzing
the resulting data associated with each feature individually, we have verified
(where applicable) that our measurement of the SMICA map's $p$-values
(single-tail probabilities) relative to the simulation ensembles are
consistent with previous findings.

\fig{fig:1dhists} shows a summary plot of these anomaly statistics measured
from our ensembles of 100,000 synfast (left column), and 1,000 FFP (right
column) simulations. In each panel, the grey histogram shows the distribution
of simulation measurements, which are either made based on full-sky maps, or
the cut UT78 sky, as indicated by the gray text in the lower right
corner. These data are what will be used in subsequent sections to study the
relationship between features in the context of isotropic $\Lambda$CDM. The
vertical lines show feature measurements done using the Planck public QML
$\Cl$'s (blue), pseudo-$\Cl$'s extracted from the full-sky SMICA map (green),
UT78 cut-sky pseudo-$\Cl$'s for the SMICA map (orange), and the theoretical
expectation based on Planck's best-fit parameter values (dashed black).  Note
that the two statistics that depend on the phases ($\SQO$ and $\ALV$) do not
have a corresponding measurement from either the published QML or the best-fit
theory $\Cl$, as they cannot be related to the angular power spectrum. The
QML and theory values for $\svar$ are computed using \eq{eq:s16expect}. The
$p$-values for these measurements are shown in the same color on the
right-hand side of each panel, with an arrow indicating which measurement we
think is most relevant for the feature in question. These choices and the
findings for each quantity are discussed in detail above.

To put the $p$-values in context, the 1, 2, and 3$\sigma$ error bars for a
normal distribution correspond to single tail probabilities of $p=16\%$,
$2.2\%$ , and $0.014\%$, respectively. Thus, of the features we study, the
SMICA values for $\Shalf$, $\quadrupole$, $\svar$, $\SQO$, and $\ALV$ are
$2-3\sigma$ unlikely compared to our simulations, while $\Rparity$ and $\ctpi$
are between $1\sigma$ and $2\sigma$, and $\octopole$ is not in tension. For
the most part our measured $p$-values are consistent with previous
findings. Where they are not, we can explain the discrepancies in terms of the
power spectrum measurement technique (for $\Rparity$) or the initial map
resolution (for $\ALV$).

We can also use the results in \fig{fig:1dhists} to make some general
observations about the impact of different measurement choices. Outside of
small differences which are within the reasonable range of sampling error, the
one-dimensional $p$-values comparing measurements of the real sky measurements relative to synfast simulations are in good agreement with those comparing the real sky to  FFP simulations. We additionally note that in general the Planck
public QML $\Cl$'s give results that are very similar to full-sky $\Cl$
measurements. The fact that the cut-sky pseudo-$\Cl$ SMICA measurements have
comparably lower $\Shalf$, $\quadrupole$, $\octopole$, and $\svar$ are
consistent with previous studies which have found that the observed lack of
power in the CMB sky is more severe in regions further from the galactic mask.


\section{Results: Anomaly covariances}\label{sec:results}

We are now ready to tackle the main goal of this paper and study the 
relationships between the large-angle CMB temperature map properties which have 
been examined individually above. The features are: the integrated power of 
temperature fluctuations at angles $\theta>60^{\circ}$ ($S_{1/2}$), the 
quadrupole amplitude ($C_2$), the octopole amplitude ($C_3$), the variance of 
the temperature map evaluated at
resolution $\nside=16$ ($\svar$), the parity statistic $R$ with
maximum multipole of $\ell=27$ ($R_{27}$), the angular power spectrum at $180\degr$
($C(\pi)$),  the quadrupole-octopole alignment ($\SQO$), and the amplitude of the hemispherical power asymmetry ($\ALV$). Using our measurements of these quantities from our synfast and FFP simulations, we will determine their covariances in order to build an understanding of how they are related under the assumption of isotropic $\Lambda$CDM.  We will do so in three stages, first describing the relationship between pairs of features measured from the synfast simulations, then comparing the covariance matrices for the synfast and FFP simulations, and finally, further exploring the covariance structure by performing a principal component analysis.

\begin{figure*}
\centering
\includegraphics[width=\textwidth]{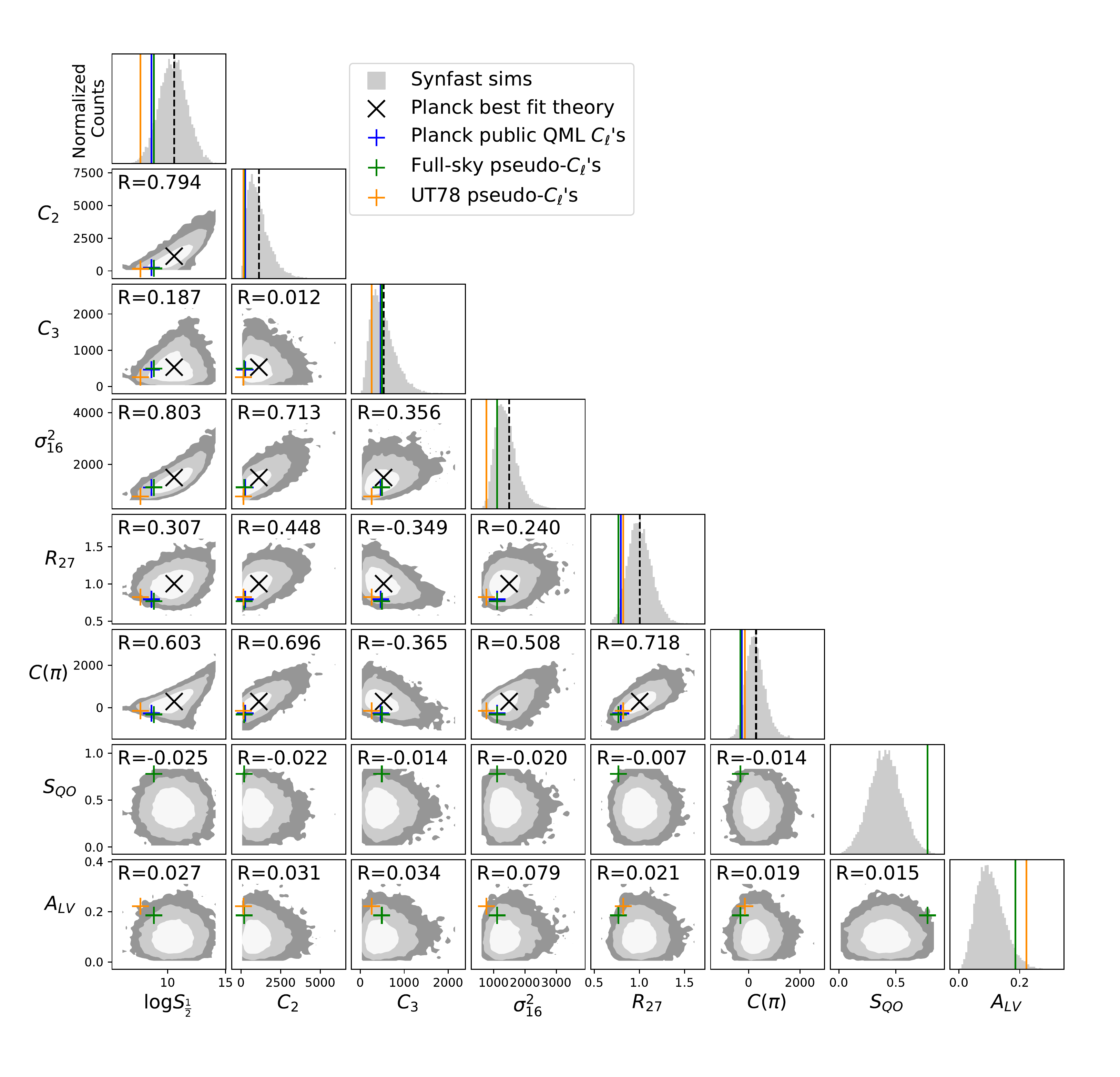}
\caption{Relationships between large-angle CMB features (see
  \tab{table:anomalies_list} for descriptions of these quantities). Gray
  contours show the 1, 2, and $3\sigma$ confidence regions based on
  measurements from our ensemble of 100,000 synfast simulations, using the
  same measurement choices that produced the gray histograms in the left column of 
  \fig{fig:1dhists}.  The 1D histograms  on
  the diagonal are the same as those in the left column of
  \fig{fig:1dhists}. The marked data points for SMICA measurements and
  theoretical expectations are equivalent to the vertical lines in
  \fig{fig:1dhists}. Note that the statistics based on the phases
  of the $\alm$, $\SQO$ and $\ALV$, do not have the corresponding
  theoretical expectations because they cannot be computed analytically from $\Cl$.}
\label{fig:triangle}
\end{figure*}

We begin by inspecting how our fiducial set of 100,000 synfast simulations are 
distributed in the eight-dimensional space defined by the parameters $\Shalf$, 
$\quadrupole$, $\octopole$, $\svar$, $\Rparity$, $\ctpi$, $\SQO$ and $\ALV$ introduced in \sect{sec:anomalydef}. 
\fig{fig:triangle} shows the relationships between pairs of those quantities. The 
diagonal panels display the same one-dimensional statistical information as the 
left column of \fig{fig:1dhists}, with the gray histograms showing the 
distribution from 
simulations and the vertical lines showing the measurements of the SMICA map 
and theoretical expectations based on the Planck's best-fit cosmological 
parameters.

In the off-diagonal panels, the grey contours  indicate the 1, 2, and $3\sigma$ 
confidence regions based on simulation data. The contour locations are 
determined for each panel as follows. First, we use simulation data to make a 
two-dimensional histogram with fifty bins along each axis. Next, we smooth the histogram  using  a Gaussian filter with a width corresponding to one bin. The 
smoothed histogram is then thresholded at constant-count 
(constant-probability) surfaces so that 68\% of the input realizations fall 
inside the $1\sigma$ contours, etc. The number at the top of each panel displays the 
correlation coefficient $R$ 
of the two quantities shown, computed 
based on the simulation samples. Measurements of the observed SMICA map and 
theory predictions are shown using colored crosses.

Examining \fig{fig:triangle}, we can make a few general
observations. First, there is structure and notable covariances  in the
relationships between most of the features that depend only on two-point
functions, but not between $\quadrupole$ and $\octopole$, nor between the
$\alm$-phase-dependent quantities ($\SQO$ and $\ALV$) and any of the other
quantities. This is expected given the isotropic $\Lambda$CDM model used to
generate the simulations.  We additionally note that the distribution of
simulation points in this eight-dimensional space is decidedly non-Gaussian;
this is due to the asymmetric limits on the quantities measured,
as well as the (in some cases) nonlinear dependence of quantities on $\Cl$.

More specifically, the covariances of the two-point-function-based 
quantities can be understood by how they depend on the power spectrum 
components $\Cl$. In isotropic $\Lambda$CDM, we expect the power at different 
multipoles $\ell$ to be independent. Correspondingly we see little 
covariance between $\quadrupole$ and $\octopole$. The positive correlation 
between $\svar$ and either of these amplitudes is straightforward,  given 
\eq{eq:s16expect}: all else being equal, adding power to low $\ell$ increases the variance at large scales. Similarly, increasing 
$\quadrupole$ adds to even-$\ell$ power and increasing $\octopole$ adds to 
odd-$\ell$ power, so we expect and see that the  parity measure 
$\Rparity$ to be positively and negatively correlated with $\quadrupole$ and 
$\octopole$, respectively.

Looking at $\ctpi$ allows us to clarify the relationship between the parity
properties and $\Shalf$.  We note that, given the parity properties of
spherical harmonics and referencing \eq{eq:ctpidef}, the contributions from
even-$\ell$ modes to $\ctpi$ will be $\propto (2\ell+1)\Cl$, while odd-$\ell$
contributions are $\propto -(2\ell+1)\Cl$. Thus $\ctpi$ is effectively another
way of characterizing the parity properties of the large-angle
CMB. Accordingly we see that $\ctpi$ has a strong positive correlation with
$\Rparity$ and with $\quadrupole$ and a somewhat weaker negative correlation
with $\octopole$.  Since $\ctpi$ is the measurement of the angular correlation
at $\theta=180^{\circ}$, small $\Shalf$ values require that $|\ctpi|$ be close
to zero. The triangular structure of the contours in the
$\ctpi$-$\Shalf$ plane reflect this behavior. (The fact that the triangle is
asymmetric about $\ctpi=0$ can be understood in terms of the fact that the
simulations are based on the $\Lambda$CDM Planck best-fit power spectrum: it has $\ctpi>0$, because
the
quadrupole $\ell=2$ mode generates a dominant positive term in
\eq{eq:ctpidef}.)  Accordingly, the $\Shalf$-$\Rparity$ contours show an echo
of that triangular structure. This provides an intuitive way to
understand the result derived analytically in Ref.~\cite{Kim:2010st}.

We note that the panel showing the cross correlation between $\quadrupole$ and
$\log{\Shalf}$ is comparable to that studied in Ref.~\cite{Knight:2017kvk}, though we
find a looser relation between the two quantities.  The reason for this difference is that we mix masking choices for our
fiducial synfast map statistics, using full-sky measurements for the
$\quadrupole$ and cut-sky measurements for $\Shalf$. We verify that if we
either measure both quantities from full-sky maps or both on cut-sky maps, the shape of our
contours closely resemble the distribution in Ref.~\cite{Knight:2017kvk}. (Ref.~\cite{Knight:2017kvk}
 uses UT78 cut-sky measurements of both features.)

The results from FFP simulations are visually similar to those in
\fig{fig:triangle}, so  we do not show the scatter plot for that
ensemble. Instead, below we  examine the quantitative difference
  between the feature covariances based on the synfast and FFP sets of simulations.

\subsection{Covariance structure comparison}\label{sec:anomalycov}

Here we measure the covariance between large-angle features measured from our 
simulated CMB temperature maps, and compare the covariances extracted from 
the synfast and FFP simulations. By making this comparison, we can gauge 
whether survey properties modeled in the FFP but not in the synfast 
simulations affect the relationship between the features studied. This in turn 
can potentially provide insight into whether those survey properties 
influence observed anomalies in the SMICA map --- though the fact that no one 
has yet found a convincing systematics-based explanation for any of them makes this unlikely. 

\begin{figure*}
\centering
\includegraphics[width=.54\textwidth]{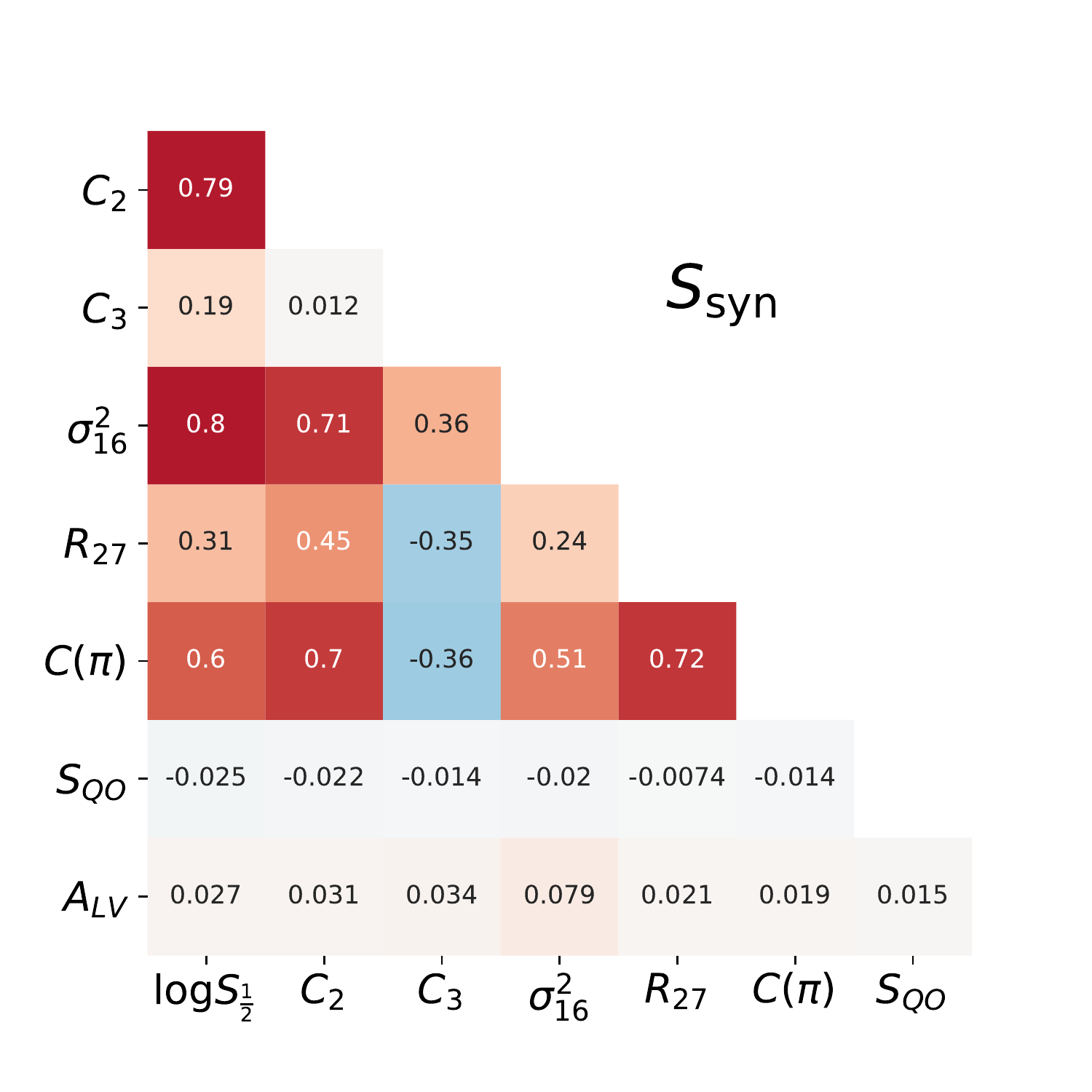}
\includegraphics[width=.44\textwidth]{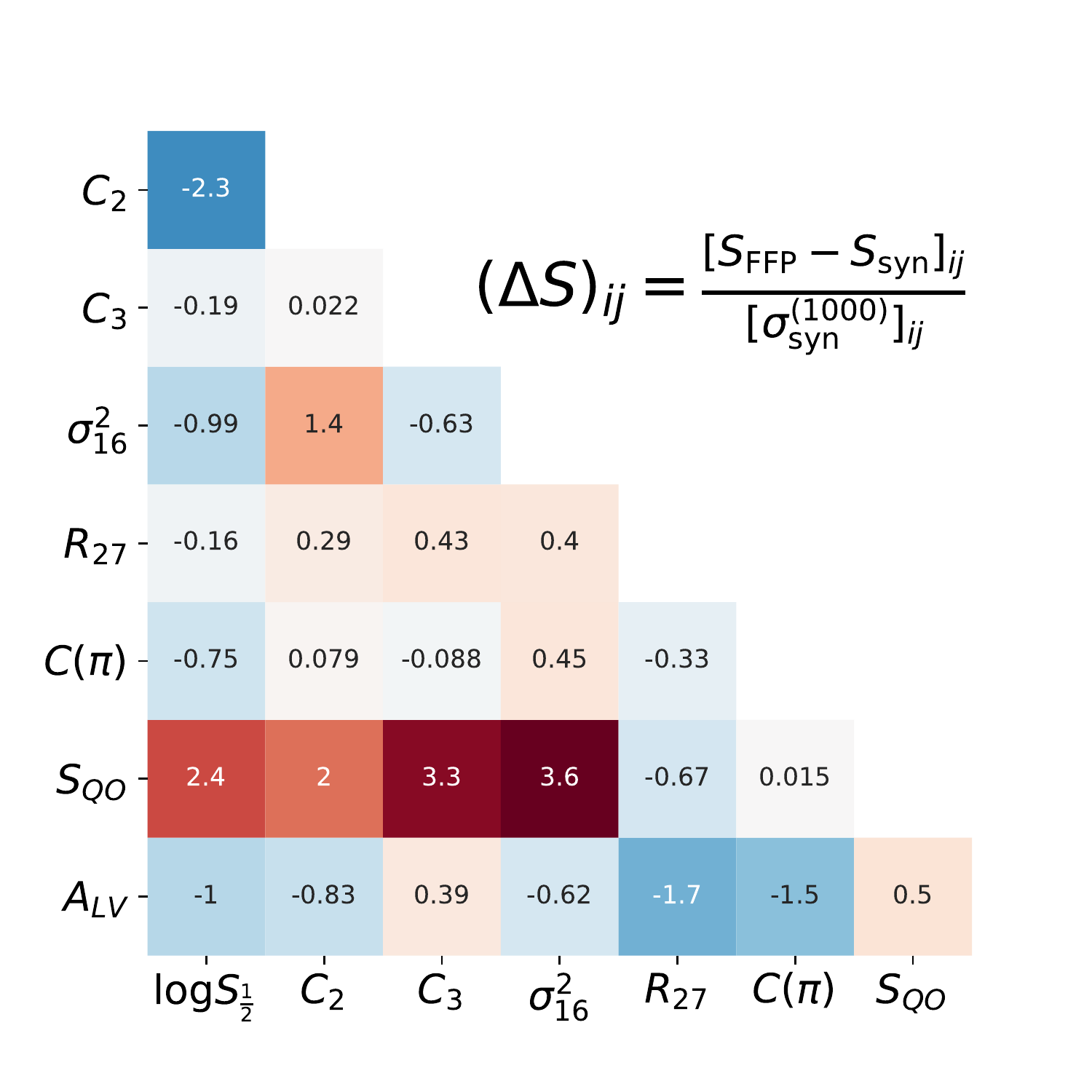}
\caption{ Left: The feature covariance matrix $S_{\rm syn}$  measured from the synfast simulation ensemble. Right:  The difference $\Delta S$ (given in \eq{eq:DeltaS}), between the covariances measured from the FFP and synfast simulations in units of its sampling error $\sigma_{\rm syn}^{(1000)}$  estimated from sets 1,000 synfast realizations (defined in \eq{eq:covsamplevar}).}
\label{fig:corrcoefs}
\end{figure*}

For a given ensemble of $n$ simulations, we represent each realization as a $d$-dimensional vector $\vec{x}$, where $d=8$ is the number of large-angle quantities measured ($\Shalf$, $\quadrupole$, $\octopole$, etc.). Before measuring the covariance matrix, we center and normalize the data so the $j$th vector component of realization $i$ becomes
\be\label{eq:preprocess_forcov}
\tilde x_i^{(j)} = ({x}_i^{(j)} - \bar{x}^{(j)})/\sigma^{(j)}, 
\ee
where $\bar{x}^{(j)}$  and $\sigma^{(j)}$ are the mean and standard deviation over realizations of the $j$th quantity being measured, respectively. This ensures that the covariance structure is not dominated by differences in the characteristic size of some of the quantities we study (e.g. $\svar$ compared to $\ALV$). It will also mean that the covariance we measure will be equivalent to  the correlation coefficients appearing in the subplots of \fig{fig:corrcoefs}.

Once the data are preprocessed, we define a $d\times n$-dimensional matrix $X =  (\tilde{\vec{x}}_1,\tilde{\vec{x}}_2,\dots,\tilde{\vec{x}}_n)$, where each column corresponds to one realization. This allows us to concisely write the measured $d\times d$-dimensional feature covariance matrix as
\be\label{eq:covdef}
  S = \frac{1}{n}\sum_{i=1}^n \tilde{\vec{x}}_i \otimes \tilde{\vec{x}}_i = \frac{1}{n}XX^T.
\ee
We show the covariance matrix $S_{\rm syn}$ for our ensemble of 100,000 synfast simulations  on the left side of \fig{fig:corrcoefs}.

We would like to study how the covariance matrix derived from the ensemble of 1,000 FFP simulations, $S_{\rm FFP}$, differs from from that measured from our 100,000 synfast simulations, $S_{\rm syn}$. To make that comparison meaningful, we must ensure that differences we see are not an artifact of sample variance due to the smaller number of FFP simulations.  On the right side of \fig{fig:corrcoefs} we therefore show the relative difference for covariance matrix entries $S_{ij}$,
\begin{equation}\label{eq:DeltaS}
\left(\Delta S\right)_{ij} =   \frac{\left[(S_{\rm FFP})_{ij} - (S_{\rm syn})_{ij}\right]}{[\sigma^{(1000)}_{\rm syn}]_{ij}},
\end{equation}
where the denominator $[\sigma^{(1000)}_{\rm syn}]_{ij}$ is sampling error for
when $S_{ij}$ is measured from a set of 1,000 synfast simulations. 

We estimate $\sigma^{(1000)}_{\rm syn}$ based on $N=100$ subdivisions of the
100,000 synfast simulations. This allows us to measure the covariance matrix
$S_{\rm syn}^{(1000,\alpha)}$  for each subsample
$\alpha\in\{1,\dots,N\}$. For each entry $ij$ of the matrix, we can then
compute the mean over subsamples $[\bar{S}_{\rm syn}^{(1000)}]_{ij}$, as well
as  the sample variance, 
\begin{equation}\label{eq:covsamplevar}
  [\sigma^{(1000)}_{\rm syn}]_{ij}^2 = \frac{1}{N-1}\sum_{\alpha=1}^{N}\left([S_{\rm syn}^{(1000,\alpha)}]_{ij} - (\bar{S}_{\rm syn}^{(1000)})_{ij}\right)^2.
\end{equation}
Thus, assuming the errors on the covariance matrix entries are Gaussian, the values plotted in the right panel of \fig{fig:corrcoefs} show the difference between the FFP and synfast covariances in units of their 1000-synfast-realization-based standard deviation. Plots of the absolute difference $S_{\rm FFP} - S_{\rm syn}$ and the sampling error $\sigma^{(1000)}_{\rm syn}$ are shown for completeness in Appendix~\ref{sec:ffpcov}.

We see that there are several moderately significant differences between the
feature covariances derived using the FFP and synfast simulations. The most
prominent of these are between the quadrupole-octopole alignment $\SQO$ and
$\log{\Shalf}$, $\quadrupole$, $\octopole$, and $\svar$, with differences
ranging from (2--3.6)$\sigma$. There is also a $2.3\sigma$ difference in
the $\Shalf$--$\quadrupole$ entry, and several other  less significant
differences  in the range (1--2)-$\sigma$. 

Noting that the largest $\Delta S$ entries  involve the quadrupole, we
hypothesized that these differences might be driven by the kinematic
quadrupole, which is partially simulated in the FFP maps but not in the
synfast simulations. To test this idea, we created an alternative version
of the synfast ensemble where the Doppler quadrupole (DQ) correction  $a_{2m}^{\rm DQ}$
given in Table~3 of Ref.~\cite{Copi:2013jna} is added to each
map\footnote{This DQ correction is slightly different than that included
    in the FFP simulations, which model only the residual frequency-dependent
    portion of the kinematic quadrupole that is not removed during the Planck map
    processing. }.
When we use this synfast+DQ ensemble to reproduce \fig{fig:corrcoefs}
there are no significant changes in the feature covariance matrix or its differences from
the FFP feature covariance. If the  $S_{\rm FFP}-S_{\rm syn}$ differences were mainly
  driven by the kinematic quadrupole present in the FFP maps, we would expect that
  adding a DQ correction to the synfast simulations would significantly change the
  structure of $\Delta S$. Because it does not, we conclude that modeled foregrounds or
survey properties other than the kinematic quadrupole 
are driving the
differences between $S_{\rm FFP}$ and $S_{\rm syn}$.

\subsection{Principal component analysis}\label{sec:pca}

We next use a principal component analysis (PCA) to investigate whether
large-angle CMB anomalies can be reduced to a few fundamental ``building
blocks'' --- features, or combinations thereof, which explain the ways that
the observed CMB sky is unusual compared to our ensembles of simulations.
Ref.~\cite{Schwarz:2015cma} conjectured that there are three
such building blocks in the CMB maps observed by WMAP and Planck: missing
large-angle power, alignments between the low multipoles, and dipolar
modulation of the CMB (which is roughly equivalent to the hemispherical
asymmetry studied in this paper). We now have an opportunity to quantitatively test
this conjecture by using our simulation measurements. By finding the simulation data's  principal
components (PCs) in the eight-dimensional feature space we consider, we can determine which linear combinations of features explain most of the
covariance structure discussed in \sect{sec:anomalycov}. It is our hope that 
studying the position of the SMICA in this PC basis will allow us to further characterize the large-angle properties of the observed CMB temperature map.

PCA is a dimensionality reduction technique which works by identifying the
directions in a $d$-dimensional parameter space along which a set of data
points have the maximum variance.  The principal components are defined sequentially: the first PC
corresponds to the direction in which our simulation realizations have the
most variance; the second PC corresponds to the direction of maximum variance
after the components of the data in the direction of PC 1 are projected out;
and so on. In practice, the PCs are obtained by finding the eigenvectors of
the data's covariance matrix. Therefore, in our analysis, we determine the
eigenvectors of the covariance matrices $S$ derived above in \eq{eq:covdef}  to obtain PCs which are unit-length vectors in the basis defined by the  quantities
$\Shalf$, $\quadrupole$, $\octopole$, $\svar$, $\Rparity$, $\ctpi$, $\SQO$,
and $\ALV$.  The first principal component is the eigenvector with the largest
eigenvalue, the second PC has the second largest eigenvalue, and so on.

PCA works as a dimensionality reduction technique because we can capture much of the information about the input data's variance by projecting it onto the first $d'\leq d$ PC directions. Heuristically, the fraction of the information that is retained in this projection is equal to ratio between the sum of the first $d'$ covariance matrix eigenvalues to the sum of all $d$ eigenvalues. To quantify the relative importance of the various PCs, we  adopt the complement of this quantity, the
fractional residual variance (FRV), the fraction of the
variance that is \textit{not} captured by the first $d'$ PCs. It is given by the expression
\begin{equation}
{\rm FRV} \equiv 1- \frac{\sum_{i=1}^{d'}\lambda_i}{\sum_{i=1}^{d}\lambda_i},
\end{equation}
where the eigenvalues have been ordered so that $\lambda_i\geq\lambda_{i+1}$.

\begin{figure*}
\centering
\includegraphics[width=.5\textwidth]{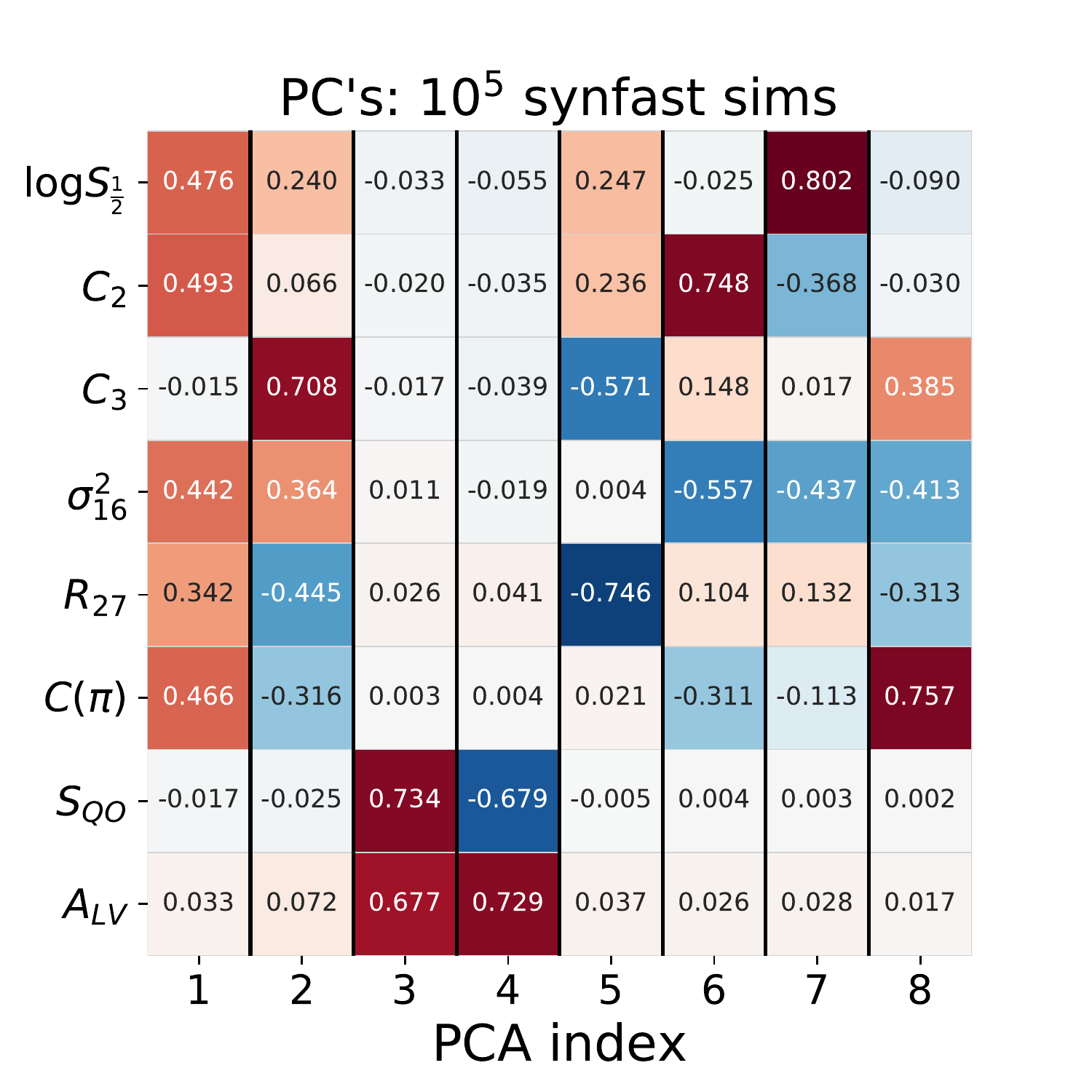}
\includegraphics[width=.48\textwidth]{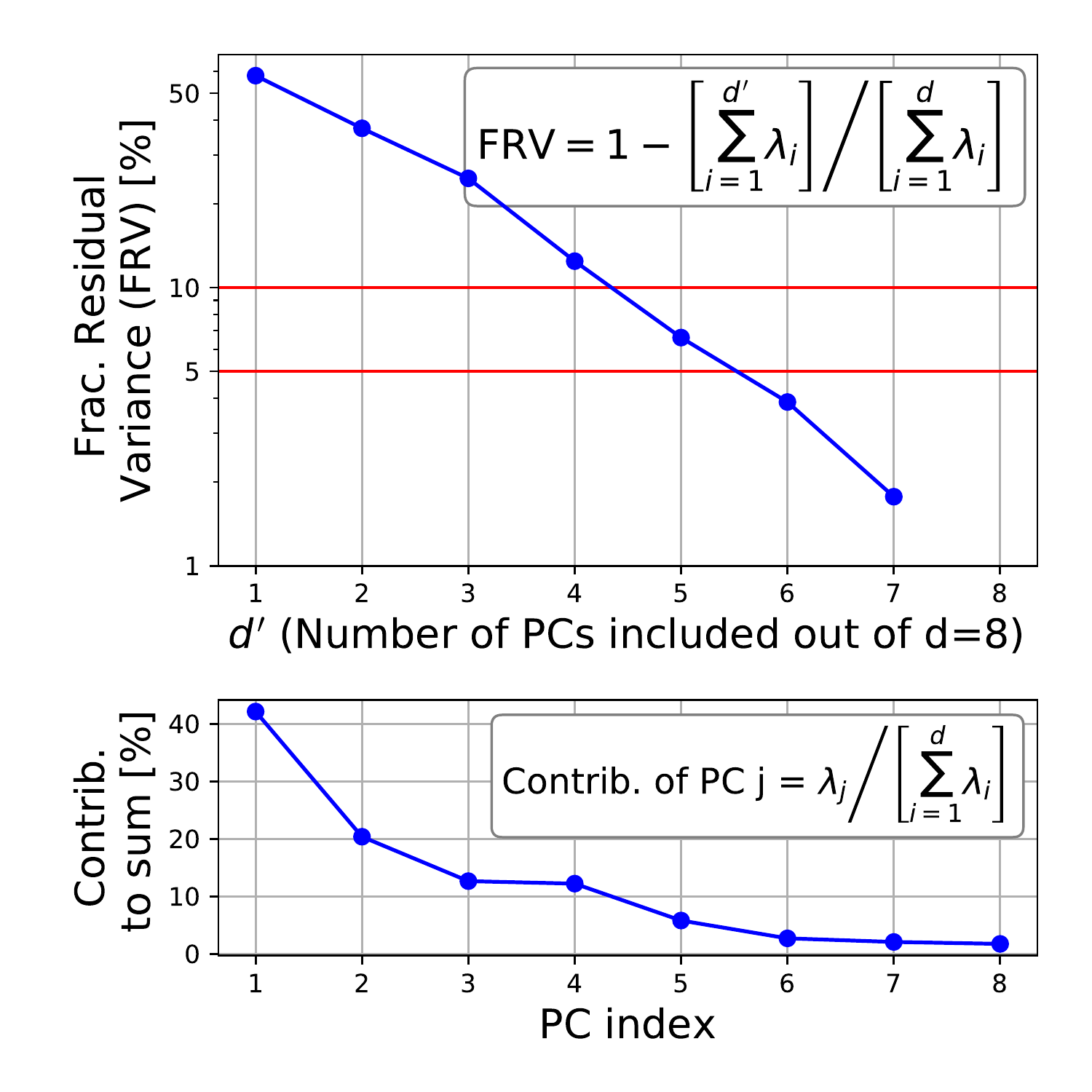}
\caption{Left: Eigenvectors of the anomaly covariance for our fiducial set of
  $10^5$ synfast simulations. Each column is one PC; the one with index 1
  points in the direction of the data's maximum variance. The rows indicate
  the features' contributions to each PC.  Right: Fractional residual variance
  for these PCs and the eigenvalues associated with each PC as a fraction of
  the total sum of all eigenvalues.}
\label{fig:pca_frv_eigvecs}
\end{figure*}

Fig~\ref{fig:pca_frv_eigvecs} shows the properties of the PCs derived from our ensemble of 100,000 synfast simulations. The left panel show PCs (the eigenvectors of $S_{\rm syn}$), with each column corresponding to one PC. The rows correspond to contributions of each of the original eight quantities to the PCs. The right panel shows  the
fractional residual variance as a function of the number of PCs retained as well as the individual contribution of each PC to its sum.

Studying the eigenvectors themselves, we find that interpretation of the first
four PCs is fairly straightforward because they can be associated with input
quantities which are largely independent of one another. Together, they capture about 90\% of the simulations' variation in the space of our eight measured features. 

The first PC, which captures 42\% of the simulations' variance, roughly corresponds to missing large-angle correlations, as it is
dominated by $\quadrupole$, $\Shalf$, $\svar$ and $C(\pi)$ (and, to an extent,
$R_{27}$). These particular features are positively correlated, with
correlation coefficients $R$ varying between about 0.5 and 0.8 (see
\fig{fig:triangle}), so it is mathematically expected that they would form a
principal component whose eigenvector components have same signs and
comparable amplitudes, as we observe in PC 1. One can understand the
relationship between these features by noting how their quantities will change
if we change the quadrupole amplitude, all else being equal. Raising
$\quadrupole$ will increase large-angle power, and will increase the relative
power in even-parity modes compared to odd-parity modes, so it make sense that
$\Shalf$, $\svar$, $\ctpi$, and $\Rparity$ will all increase.  Thus, one
interpretation of the first PC is that it picks out direction in our feature
space similar to that associated with variations in the quadrupole amplitude.

Next, PC 2 accounts for 20\% of the simulations' variance and is dominated by the octopole. Given the correlations of $\octopole$
and the other statistics in \fig{fig:triangle}, it is also unsurprising that
PC 2 receives moderate contributions from $\Shalf$ and $\svar$ with the same
sign as $\octopole$, and from $ R_{27}$ and $C(\pi)$ with the opposite
sign. As with PC 1, we can understand this in terms of how how other
quantities will respond if we raise or lower $\octopole$ without changing
power at other multipoles. More octopole power will generally add to
large-angle power, increasing $\Shalf$ and $\svar$, but specifically through
odd parity contributions which correspond to lower $\Rparity$ and $\ctpi$.

The third and fourth PCs are associated with the $\alm$-phase dependent
quantities. PC 3 is associated with the sum of the quadrupole-octopole alignment
$\SQO$ and the hemispherical asymmetry statistic $\ALV$, and captures 13\% of the simulations' variance, while PC 4 is associated with their difference and captures 12\% of the variance.  We note that because their associated covariance
eigenvalues $\lambda_3$ and $\lambda_4$ are nearly equal,  the ordering of
PC 3 and PC 4 is somewhat arbitrary. This reflects the fact that the
correlation between $\SQO$ and $\ALV$ is very small, and means that using PC 3
and PC 4 together is basically equivalent to defining two unit vectors in
the $\ALV$ and $\SQO$ directions.

The structure of the fifth through eighth PC, which account for the remaining 10\% of the simulations' variance, resists simple interpretation. This is because
they are determined by the relationships between the non-independent
quantities, after the variation of the data in the direction of the first four
PCs (roughly $\quadrupole$, $\octopole$, $\SQO$ and $\ALV$) are projected
out. One could infer, for example, that because PC 5 has $\quadrupole$ and
$\octopole$ components with different signs, that it might capture some
information about whether the power from the quadrupole and octopole cancel
one another, but this is far from clear. PCs 6-8 all have small eigenvalues
with $\lambda_6\sim\lambda_7\sim\lambda_8$, so their order and the way that they divide up whatever degrees of freedom are left after the first five PCs are removed are somewhat
arbitrary.

We also performed a PCA on the FFP simulation data. The results are very
similar to those for the synfast simulations, and any differences that exist mainly just reflect
the differences between the structure of the synfast and FFP covariance
matrices discussed in \sect{sec:anomalycov}. Given this, we do not show the FFP-based
PCs.

\begin{figure*}
\centering
\includegraphics[width=\textwidth]{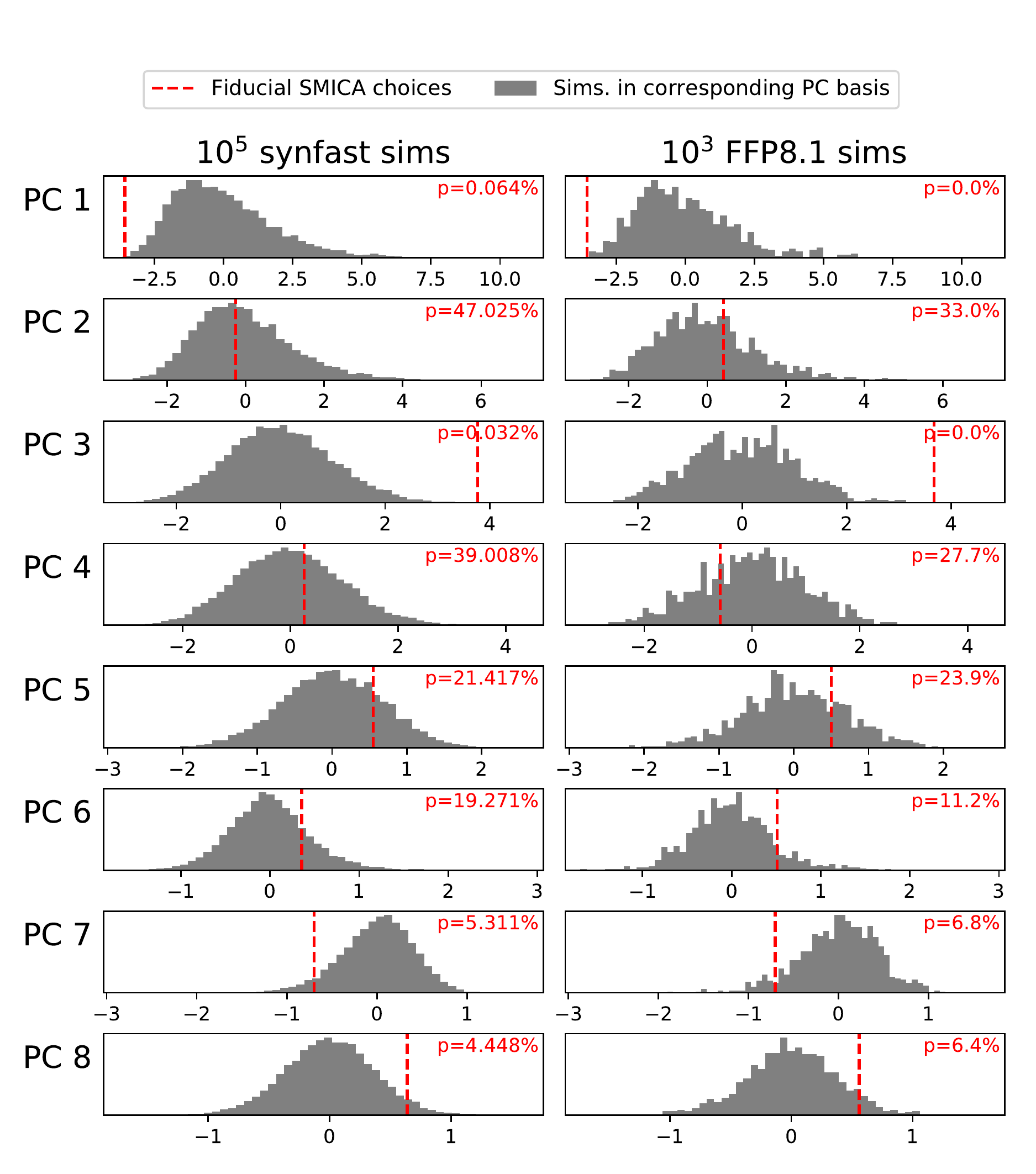}
\caption{Summary of feature statistics projected into our PCA basis. The row
  labels indicate the basis vectors of the PC basis, which are equivalent to
  the unit-eigenvectors of the synfast (left) and FFP (right) simulations'
  feature covariance matrix. The gray histograms show the distribution of the
  components of simulation realizations in the direction of each PC. The red
  lines show the same projection of our fiducial SMICA map measurements (using
  the measurement methods whose $p$-values in \fig{fig:1dhists} are denoted by
  an arrow). The red numbers in the top right corner of each panel
    show the percentage of simulations that are more extreme than the
    corresponding SMICA measurement. }
\label{fig:pcahists}
\end{figure*}

We next calculate the probability of CMB maps projected to the PC basis. This can inform whether linear combinations of the features that tend to
``come together'' in simulated skies are particularly anomalous (or not) when
observed on our CMB sky. We proceed as follows. The measurement of each simulation corresponds to a vector in our eight-dimensional
feature space. By taking the dot product of that vector with each PC,
we find its components in the new PC basis.  The
grey histograms in \fig{fig:pcahists} show the resulting distributions of the
simulations projected to the PC basis, where as before we show results from the synfast simulations in the left column, and from the FFP simulations in the right column. The red vertical lines correspond to the observed CMB sky, using the fiducial SMICA map measurements (indicated in \fig{fig:1dhists} by arrows next to their $p$-values).

\fig{fig:pcahists} shows some instructive trends. Since the first
principal component (PC 1) is a linear combination of the features that encode
the missing angular correlations (low $\quadrupole$, $\Shalf$, $\svar$ and
$C(\pi)$), it makes sense that the probability observed sky's PC 1 coefficient is low.  However the fact that
this probability is lower than that for any of the individual
features ($p$-value $=0.064\%$) further indicates that, even given the lowness of
one of its constituent statistics (for example, the quadrupole), the other
 features that make up PC 1 are still lower than expected in
$\Lambda$CDM. Next, the probability of PC 2 is not anomalous ($p=47\%$), which is unsurprising
given that it largely reflects the observed sky's rather average $\octopole$. The
PC 3 probability, however, is surprisingly high ($p=0.032\%$), which is the smallest
p-value among all PCs. The extremely high value of the statistics projected to
PC 3 comes from the fact that this principal component is largely a sum of
$\SQO$ and $\ALV$, which are both high on our sky but uncorrelated
  ($R=0.015$) in $\Lambda$CDM. Hence, PC 3 is the sum of two high-valued
statistics, and is therefore very high itself. In contrast, PC 4 is mainly a difference
between the same two high-valued statistics ($\SQO$ and $\ALV$), and so is itself
 average ($p=39\%$). The higher PCs  do not shed
significant further light on the statistical properties  of the features
we study.

In concluding this section, we caution that PCA as a method is only able to
capture linear structures in the dataset. Because the relationship between
many of the quantities we measure are nonlinear by definition, the PCAs will
therefore not capture all of the structure in the simulations' distribution.

\section{Conclusions}\label{sec:concl}
In this paper we have studied the relationships expected in $\Lambda$CDM between a set of large-angle
CMB features, with the goal of better
understanding the interdependence of large-angle temperature anomalies
observed in WMAP and Planck data.  In particular, we have studied eight
features measured via the quantities defined in \sect{sec:anomalydef}: the
integrated power of temperature fluctuations at angles $\theta>60^{\circ}$
($S_{1/2}$), the quadrupole amplitude ($C_2$), the octopole amplitude ($C_3$),
the variance of the temperature map evaluated at resolution $\nside=16$
($\svar$), the parity statistic R with maximum multipole of $\ell=27$
($R_{27}$), the angular correlation function at $180\degr$ ($C(\pi)$), the
quadrupole-octopole alignment ($\SQO$), and the amplitude of the hemispherical
asymmetry ($\ALV$). The first six of these features depend on the angular
power spectrum and quantify various aspects of isotropic angular clustering at large
scales, while the last two depend on phases of the $\alm$ and quantify
large-angle alignments and power asymmetry observed in CMB maps. In addition to several commonly studied anomalous features,  this list includes a few features which have not been individually reported to be   anomalous in the CMB data (the octopole is a notable example), but which have been flagged in previous work as potentially interesting in relation to the other  features. 

Our analysis was based on on measurements of two ensembles of $\Lambda$CDM
simulations: 100,000 noiseless Gaussian CMB temperature maps generated using
the {\tt synfast} function in {\tt healpy}, and 1000 full focal plane (FFP8.1)
simulations provided by the Planck team that contain astrophysical foregrounds
and other physical artifacts expected in the observed sky.  We began by using
these ensembles to find the probability  of each
feature in $\Lambda$CDM. This allowed us to study the impact of analysis
choices on the features' statistics and to make sure we could recover results
from previous work. We found generally excellent agreement between the
statistics measured from our two sets of synthetic maps, and
summarized the results in \fig{fig:1dhists}.

Then, selecting a fiducial set of analysis choices, in \sect{sec:results}
we used those same simulation measurements to fulfill the principal goal of
this paper by calculating the correlation between the eight features
studied. \fig{fig:triangle} shows, for the first time, a complete covariance
of the features associated with the most commonly discussed
large-angle CMB anomalies. Our results confirm and  quantify various
aspects of the features that were previously either merely conjectured or calculated in isolation. For example, the quadrupole
$\quadrupole$, the missing large-angle correlations statistic $\Shalf$, and
the variance $\svar$ are all positively correlated and largely uncorrelated to
the phase-dependent features. The phase-dependent features --- the
quadrupole-octopole alignment $\SQO$ and the hemispherical asymmetry statistic
$\ALV$ --- are uncorrelated both with each other and with all other features
studied. Less trivially, the observed low $\Shalf$ essentially guarantees (in
$\Lambda$CDM) that the configuration-space clustering amplitude at the largest
observable scale, $C(\pi)$, is very close to zero, which is in fact observed
in the data. Furthermore, we find that the covariance between $\SQO$ and several other
  features, though still low compared to the covariances between other features,
  is significantly higher when measured from FFP simulations than from
synfast. Introducing a kinetic quadrupole correction to the synfast
simulations has little impact on that difference.

We then diagonalized the measured covariance matrix to obtain the principal
components of features' expected distribution in $\Lambda$CDM. This allowed us
to quantify whether most of the simulations' variation in
our eight-dimensional feature space is retained in some smaller number of PCs.
We find that 42\% of the simulations' variance is in the direction of the
first PC, which quantifies the missing large-angle correlations and has
comparable coefficients of the same sign in $\quadrupole$, $\Shalf$, $\svar$
and $C(\pi)$. Another 20\% is the PC 2 direction, which largely lies in the
direction of $\octopole$, along with less dominant contributions from features
correlated with the octopole. The next two PCs quantify the sum and difference
of the quadrupole-octopole alignment $\SQO$ and the hemispherical asymmetry
statistic $\ALV$, capturing 13\% and 12\% of the data's variance,
respectively. These first four PCs together explain about 90\% of variation in
the space of the (eight) features.

It is important to remind ourselves that apart from the few (generally
$2$-$3\sigma$) anomalies discussed here and elsewhere, the $\Lambda$CDM model
describes most of the current cosmological observations with immense
success. Given the significant cosmic variance inherent in the largest angular
scales of the CMB, as well as the absence of concrete models that are competitive with $\Lambda$CDM, we should be wary of putting too much
weight on these anomalies as motivations for new physics.  However, given the success of $\Lambda$CDM, any
observational clues as to how to build a more fundamental description of, for example, the physics of inflation or dark energy will (initially at least) take the form of small deviations from its predictions~\cite{Scott:2018adl}. Given this, we should certainly take a careful look at
reported tensions and anomalies, making sure we understand how assumptions related to
modeling and analysis  affect their significance.

 It is in this spirit that this work contributes to the
discussion of large-angle CMB anomalies: by understanding in detail how
observed features are related in $\Lambda$CDM, we can better assess the
independent ways in which our observed CMB sky is unusual, and thus whether
they might provide clues about beyond-$\Lambda$CDM physics.  An interesting
potential avenue for future work could be to study how the covariance between the
anomalies changes when assuming underlying models that are extensions of or
alternatives to $\Lambda$CDM.

\section*{Acknowledgments}

JM has been supported by the Rackham Graduate School through a Predoctoral Fellowship. DH and SA have been supported by NASA under contract 14-ATP14-0005. DH and JM have also been supported by DOE under contract DE-FG02-95ER40899.  Many thanks to Diego Molinari for answering our questions about the Planck
parity symmetry analysis.
The analysis in this paper was done using an
Anaconda~\cite{Anaconda} (version 4.3.22) installation of Python (version
2.7.13), and made use of the packages SciPy~\cite{Jones:2001} (version
0.19.1), NumPy~\cite{Oliphant:2006} (version 1.12.1),
Matplotlib~\cite{Hunter:2007} (version 2.0.2), and Healpy (version 1.10.3),
the Python implementation of Healpix~\cite{Gorski:2004by}.
Jupyter~\cite{Perez:2015} (version 5.0.0) notebooks were used for parts of the
analysis. Additionally, some of the results in this paper have been derived
using the multipole vector code available at {\tt
  http://www.phys.cwru.edu/projects/mpvectors/}, which was developed with
support from the US Department of Energy.

\bibliography{anomalies_cov}{}

\appendix
\section{Covariance difference and sampling errors}\label{sec:ffpcov}

Here we complement the results in \fig{fig:corrcoefs} in \sect{sec:anomalycov}
to show more information about the differences in the features' covariance
matrix calculated using the synfast simulations and those using the FFP
simulations.

\begin{figure*}[h]
\centering
\includegraphics[width=.5\textwidth]{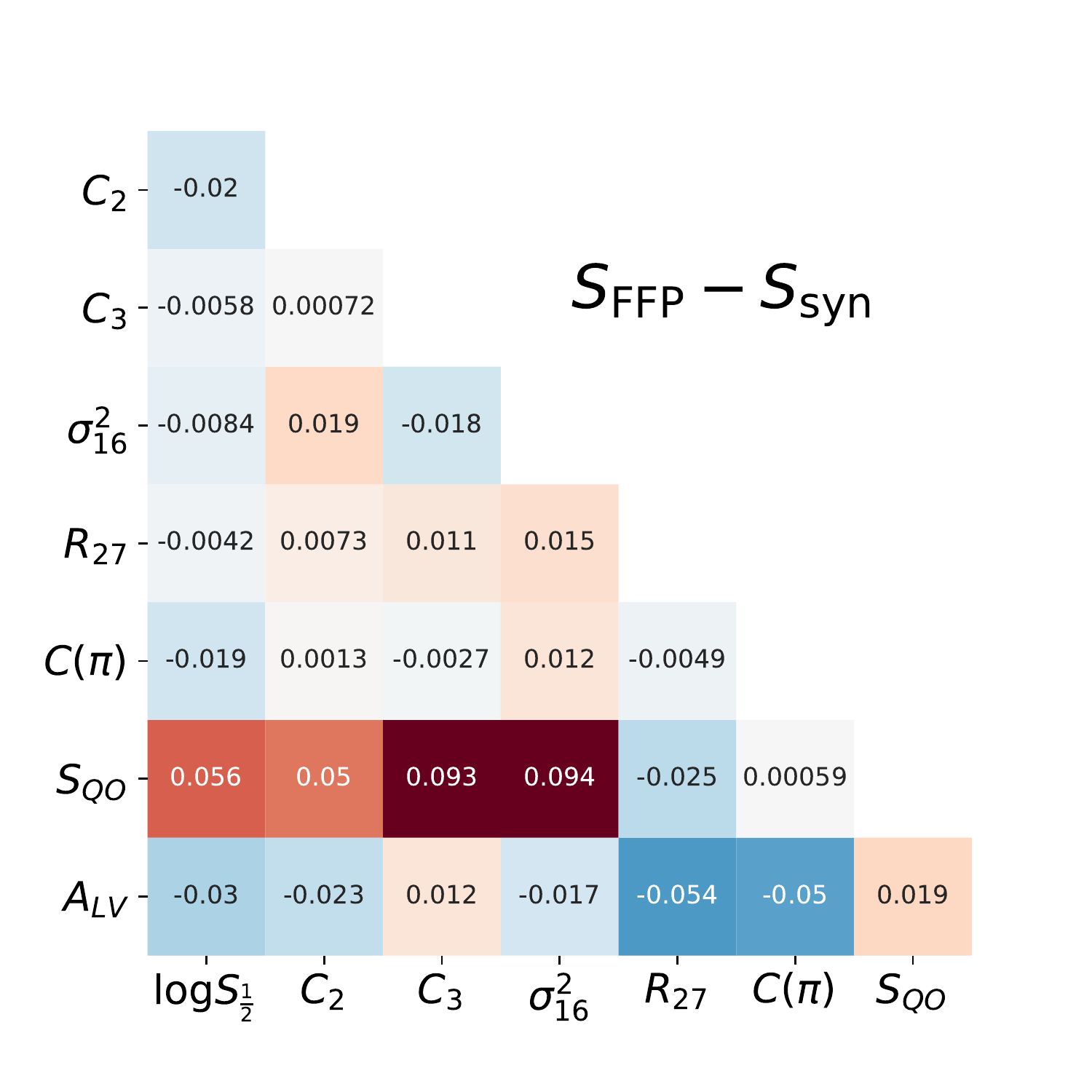}\includegraphics[width=.5\textwidth]{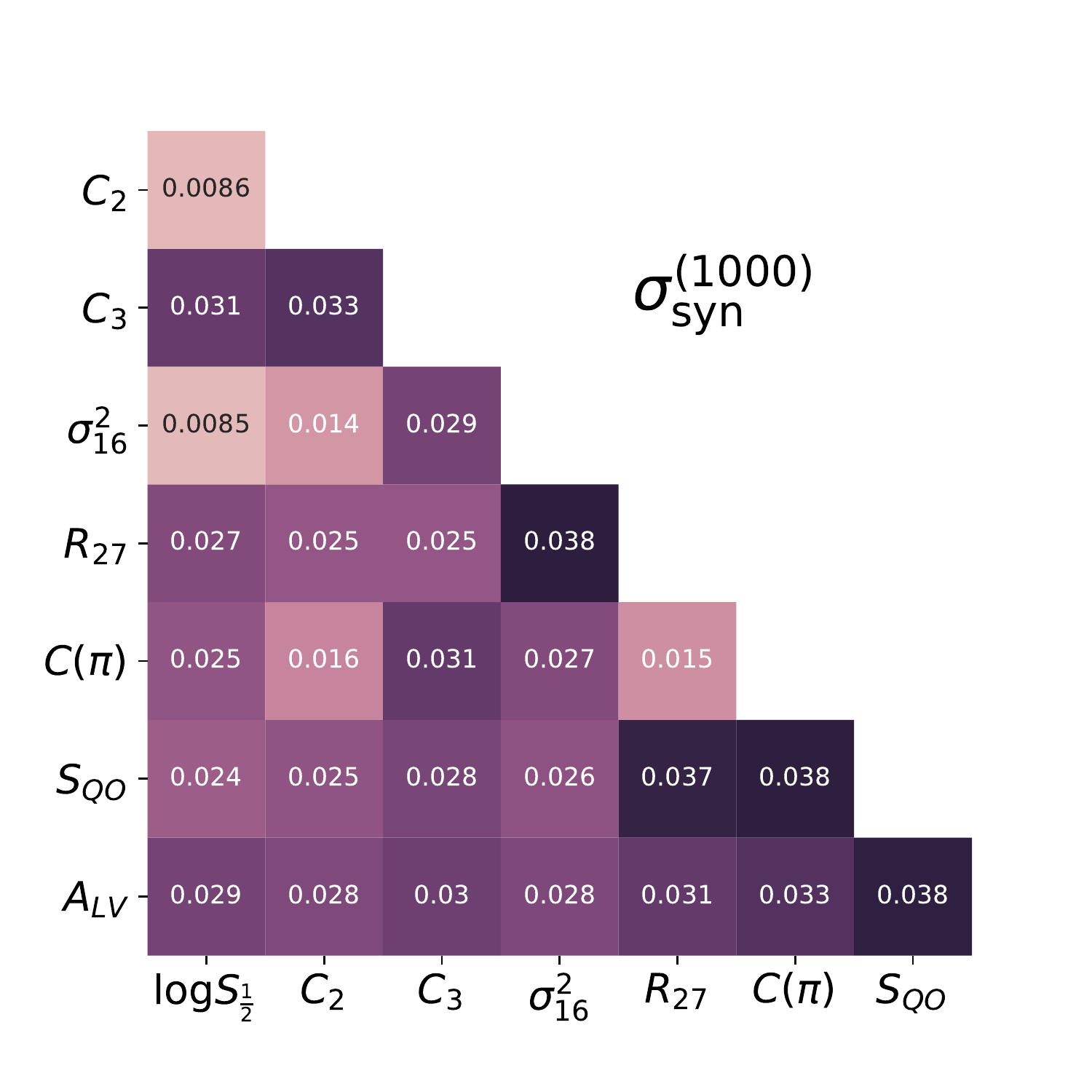}
\caption{ Left: Absolute difference between the covariance matrix measured
  from 1000 FFP simulations and from 100,000 synfast simulations.  Right:
  Sample variance error bars for covariance matrix entries measured from 100
  subsamples of the 100,000 synfast realizations, each containing 1000
  realizations.  The right panel of \fig{fig:corrcoefs} shows the ratio
  between the two panels of this figure.}
\label{fig:cov-diff-and-sigma}
\end{figure*}

The left panel of \fig{fig:cov-diff-and-sigma} shows the \textit{absolute}
differences in the coefficients calculated in the two sets of simulations. It
provides additional information to the relative differences shown in the right
panel of \fig{fig:corrcoefs} because the overall size of the correlation
coefficients (shown in the left panel of \fig{fig:corrcoefs}) varies by two
orders of magnitude.

The right panel  of \fig{fig:cov-diff-and-sigma} shows the sampling error in the correlation
coefficients in the 1000 FFP simulations, which we measure by splitting the
100,000 synfast simulations into 100 subsamples. Note that right panel of
\fig{fig:corrcoefs} shows the ratio between the two panels of  \fig{fig:cov-diff-and-sigma}.

\end{document}